\documentclass[12pt]{iopart}

\usepackage{iopams}  
\usepackage{graphicx}  

\newcommand{\beq}{\begin{equation}}
\newcommand{\eeq}{\end{equation}}
\newcommand{\bea}{\begin{eqnarray}}
\newcommand{\eea}{\end{eqnarray}}

\newcommand{\I}{\item}

\newcommand{\pr}{\mathrm{pr}}
\newcommand{\eqref}[1]{(\ref{#1})}

\newcommand{\LamchiSB}{\Lambda_{\chi {\rm SB}}}
\newcommand{\chiPT}{\chi{\rm PT}}
\newcommand{\breakdown}{\Lambda_b}

\begin{document}

\title[EFT uncertainty quantification]{A recipe for EFT uncertainty 
quantification \\ in nuclear physics}

\author{R.~J.\ Furnstahl$^{1}$, D.~R. Phillips$^{2}$, and  
    S. Wesolowski$^{1}$}

\address{$^{1}$ Department of Physics, The Ohio State University, 
         Columbus, OH 43210, USA}
\address{$^{2}$ Institute of Nuclear and Particle Physics and Department 
    of Physics and Astronomy, Ohio University, Athens, OH 45701, USA}

\ead{furnstahl.1@osu.edu, phillid1@ohio.edu, wesolowski.14@osu.edu}

\begin{abstract}
The application of effective field theory (EFT) methods to nuclear
systems provides the opportunity to rigorously estimate the uncertainties 
originating in the nuclear Hamiltonian. Yet this is just one source
of uncertainty in the observables predicted by calculations based
on nuclear EFTs. 
We discuss the goals of uncertainty quantification in such calculations and outline a recipe to obtain
statistically meaningful error bars for their predictions.
We argue that the different sources of theory error can be accounted for 
within a Bayesian framework, 
as we illustrate using a toy model.
\end{abstract}

\section{Overview} \label{sec:introduction}

Effective field theories (EFTs) are a framework by which the consequences of
known low-energy physics can be systematically computed
(see, e.g.,~\cite{Beane:2000fx}). 
The low-energy physics is specified by the choice of explicit degrees of freedom
and symmetries. Short-distance/high-energy degrees of freedom are 
accounted for in the EFT through the process of regularization and 
renormalization and a consequent power counting. The power counting is based
on an expansion in a small parameter or parameters, which are formed from ratios
between the low-energy scales (those associated with physics that is included explicitly) 
and high-energy scales (corresponding to unresolved degrees of freedom). 

We denote this expansion parameter 
as $Q$; for simplicity here we only consider $Q=p/\breakdown$ where $p$ is
a characteristic momentum and $\breakdown$ is the breakdown scale of the EFT.
At the scale $\breakdown$ the EFT fails completely because its degrees of freedom
are no longer the appropriate ones there.
If correctly and consistently implemented, power counting gives a prescription 
for calculations of observables
whose truncation error decreases systematically as higher orders in the $Q$ 
expansion are included in the EFT. 
For most nuclear physics applications this truncation error is momentum dependent.

EFTs are now common in nuclear physics.  An EFT containing only
short-range interactions (``pionless EFT") has been used for precise and accurate 
calculations in few-nucleon systems (see \cite{Epelbaum:2008ga} for a recent review).
In this case low energy corresponds to momenta $p \ll m_\pi$. An analogous 
treatment of halo nuclei (designed for 
energies of order the separation energy) is providing
insights into these distinctive systems that emerge near the 
driplines~\cite{Bertulani:2002sz}. 
At even lower energies, in even larger systems, 
a completely different EFT designed to treat the low-lying excitations in deformed nuclei
has recently been formulated~\cite{Papenbrock:2013cra}.

The dominant EFT in nuclear physics has become ``chiral EFT",
which seeks to treat multi-nucleon physics for $p \sim m_\pi$ explicitly,
and integrates out physics at scales of order 
the chiral symmetry breaking scale 
$\LamchiSB \sim m_\rho$~\cite{Epelbaum:2008ga,Machleidt:2011zz}. 
Inter-nucleon potentials and associated current operators 
derived within chiral EFT are now tools of choice for most \textit{ab initio} nuclear
structure and reaction calculations. 
In principle, chiral EFT should yield calculations in which the 
theoretical uncertainties decrease order-by-order in accord with a $(p,m_\pi)/\breakdown$ expansion. 
But uncertainty analyses in this theory
have been limited; most do not take full advantage of the uncertainty-quantification possibilities
provided by the EFT framework. 

When uncertainties have been identified for chiral EFT observables (and in many cases they still are not), this has usually been in terms of
``error bands" generated  by varying the cutoff
parameter in the EFT regulator. As long as the cutoff is kept in the vicinity 
of $\LamchiSB$, and not lowered so far as to affect the low-energy properties of the theory, 
this band provides a lower bound on the uncertainty in the EFT calculation. This is because---under these caveats---EFT Hamiltonians with different cutoffs form a ``family", differing only in their short-distance physics.
Such ``families"  of Hamiltonians can be very useful, since the spread of their predictions gives some insight into how omitted higher-order terms affect observables. However, the statistical interpretation and quantitative characteristics of the resulting
band are unclear: in some cases it seems to mirror the order-by-order convergence of the calculation, and in others it severely
underestimates the theoretical uncertainty of the calculation. 

\begin{figure}[tbh-]
 \begin{center}
 \includegraphics[width=0.36\columnwidth]{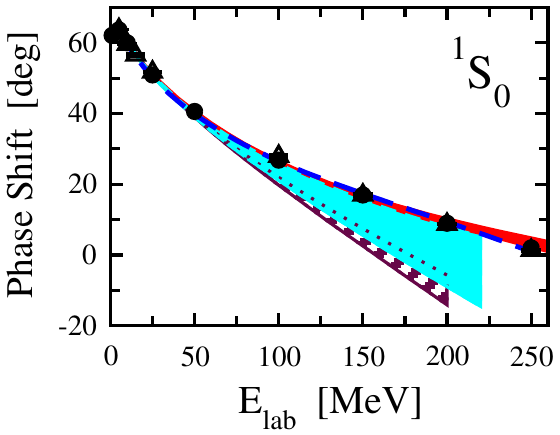}~~%
   \includegraphics[width=0.31\columnwidth]{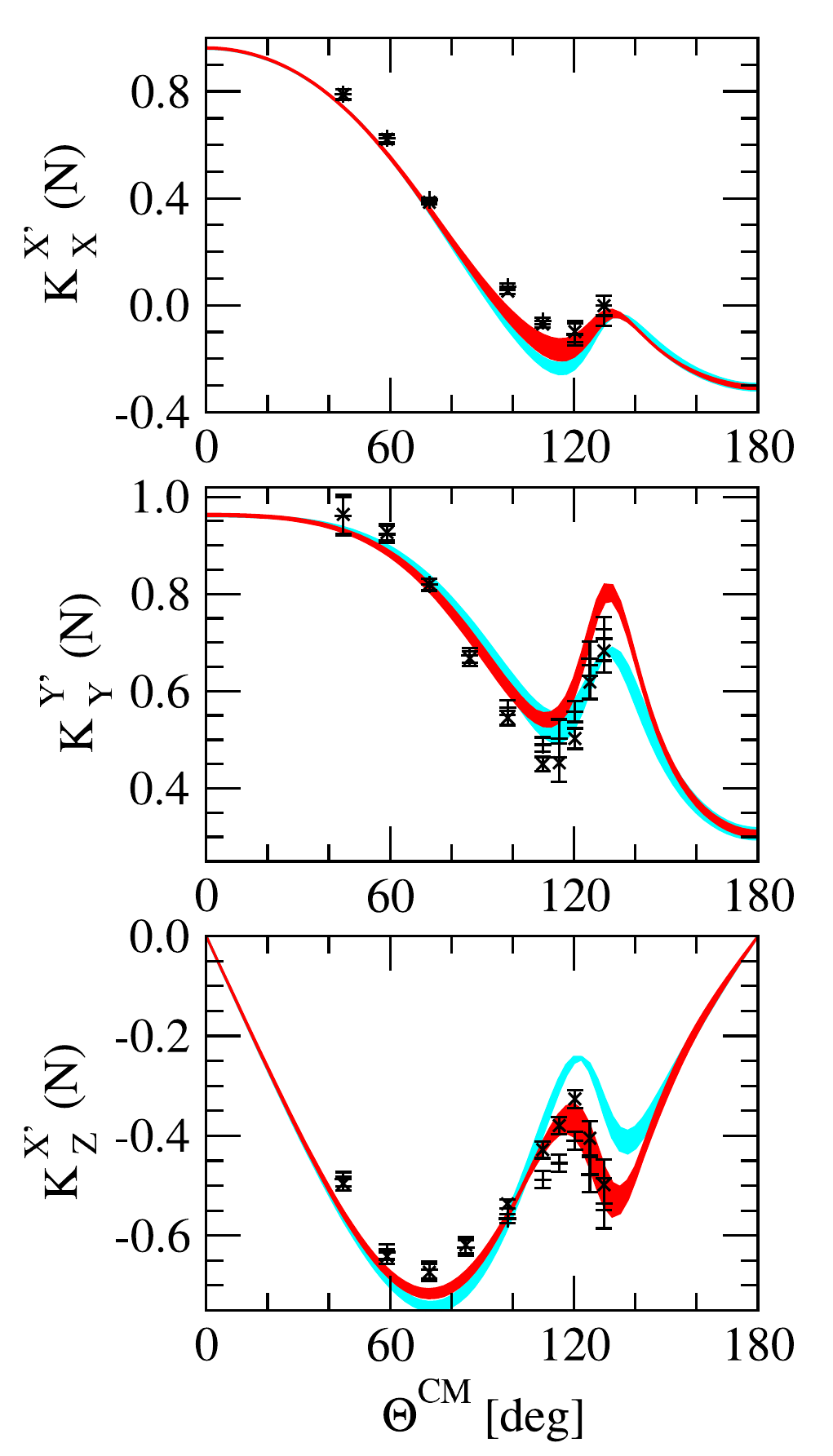}%
   \includegraphics[width=0.31\columnwidth]{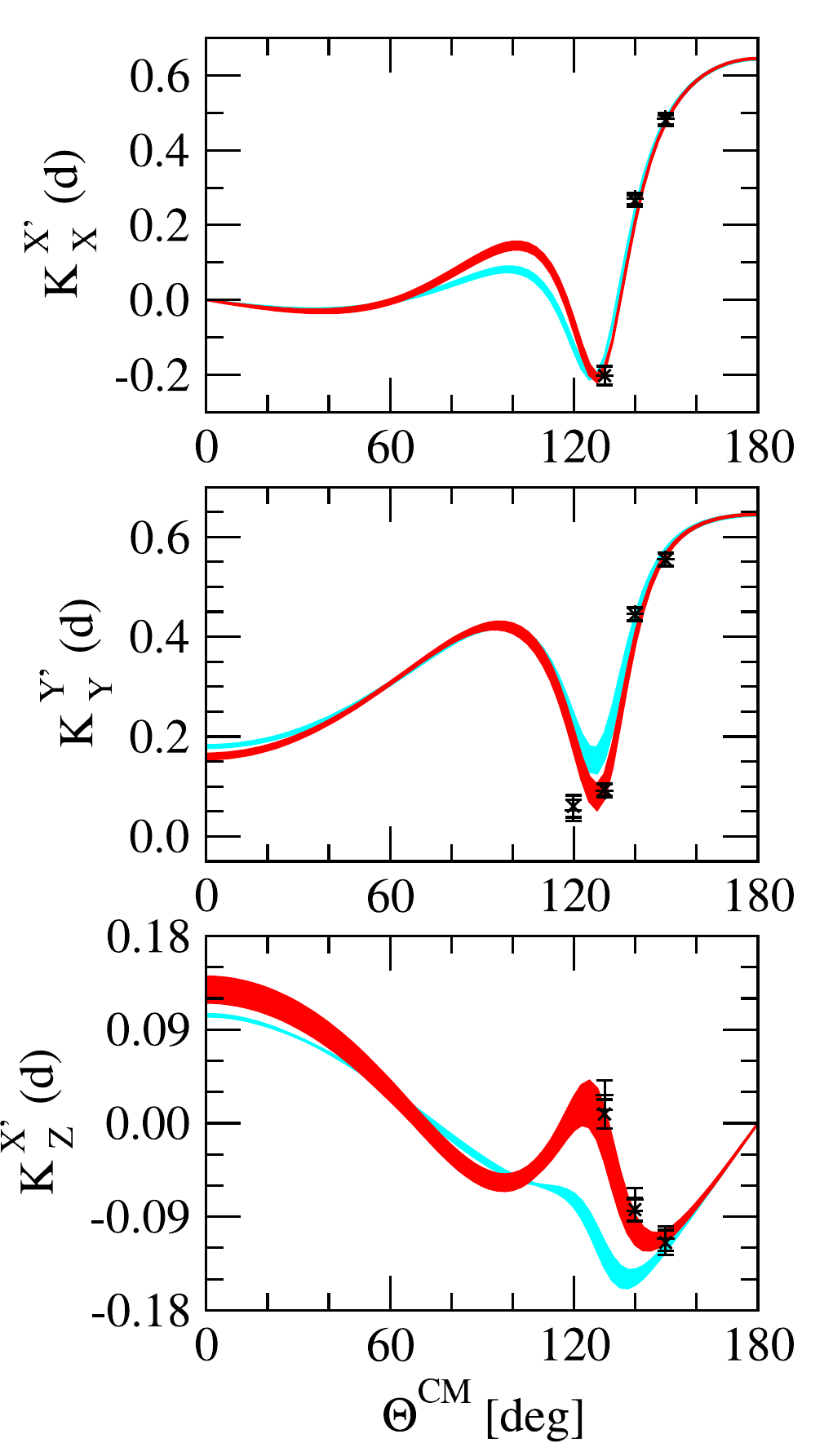}
\caption{Left: neutron-proton ${}^1$S$_0$ partial wave at NLO (hatched band), N$^2$LO (light-shaded band) and N$^3$LO (dark-shaded band) in comparison with extractions from experiment (filled circles and open triangles)~\cite{Epelbaum:2008ga}.
Right: polarization transfer coefficients in
$d(\vec{p}, \vec{p})d$ (left panel) and $d(\vec{p},\vec{d})p$ (right panel) reactions at an incoming proton laboratory energy of $22.7$ MeV~\cite{Epelbaum:2008ga}. Light (dark) 
shaded bands show the NLO (N$^2$LO) chiral EFT prediction.  Symbols represent experimental data/phase-shift analyses, for details see Ref.~\cite{Epelbaum:2008ga}.
}
 \label{fig:error_bands}
 \end{center}
\end{figure}

For example, in Fig.~\ref{fig:error_bands} the left panel shows chiral EFT calculations of the ${}^1$S$_0$ NN phase shift at NLO, N$^2$LO and N$^3$LO.  The error bands get progressively narrower, as expected for a calculation that is improving from $O(Q^2)$ to $O(Q^3)$ to $O(Q^4)$. Moreover, they overlap at each order, as they should if they accurately reflect the theoretical uncertainty. This is typical of observables that are used to fit EFT LECs at several different orders. 
In contrast, the right panel shows proton-deuteron scattering observables, which are a chiral EFT \emph{prediction}. Here the error bands do not behave in such a pleasing manner: the N$^2$LO (dark) band is frequently wider than the NLO (light) band. Should we conclude that the theory uncertainty has increased? Moreover, the shift from NLO to N$^2$LO, when sampled over the six panels, appears larger than it should be if the NLO band is interpreted as a 1$\sigma$ confidence level.
The agreement with data at N$^2$LO 
\emph{is} markedly better than at NLO: from a phenomenological point of view there is improvement.
But here, and elsewhere, the error bands obtained by scanning over this family of chiral EFT Hamiltonians do not behave in a fashion that is consistent with their oft-accorded status as the theoretical uncertainty of the calculation. 

In contrast, we seek to assign theory errors to EFT calculations that:
\begin{enumerate}
\item Reflect \emph{all} of the sources of uncertainty in the calculation. These include uncertainties in the input data (experimental errors), in the
Hamiltonian (EFT-truncation errors and regulator artifacts), and in the many-body method (e.g.\ approximations made in Hamiltonian diagonalization).
These three sources of errors may become entangled. For example, the choice of a particular regulator can affect the convergence of 
basis expansion methods. 

\item Enable a comparison between theory and experiment  in statistical terms. 
If the error bands truly reflect all sources of uncertainty, they should be 
interpretable as theory confidence intervals.
It is clear from this point of view that fine-tuning the theory to particular 
observables can be misguided, especially if the Hamiltonian and method 
uncertainties associated with the observable in question are not taken into 
account. It is dangerous to insist on reproducing a data point measured with 1\% 
precision to within that uncertainty if, in reality, 
the accuracy of the theory for that quantity is only 20\%. Such a failure to 
account for all sources of uncertainty in the theory will lead to spurious 
precision in the Hamiltonian, which may in turn lead to erroneous conclusions 
about the theory's ability to describe other observables. 

\item Provide clear guidance on the best way to extract the EFT parameters 
(the ``low-energy constants'' or LECs, see Sec.~\ref{sec:LECs}). 
In any EFT, selected data are used to fit the LECs, and once this is 
done other data can be predicted. The accuracy
of these LECs, and our knowledge of that accuracy, 
thus affects all subsequent EFT predictions. 
In chiral EFT, the data used to determine LECs might 
be NN cross sections and asymmetries, few-nucleon observables, 
energies of nuclei---or even lattice QCD calculations of these observables. 
Depending on which of these we use for LEC determination, we can 
face an interplay of all three sources of error. If NN data is used
to fix the LECs it will have more statistical power if a larger
energy range is chosen for the fit. But, as the energy range is expanded, the EFT becomes
less accurate: there is trade-off between increasing statistical weight of the data
and the 
impending failure of the EFT at the breakdown scale. 
This dilemma is not drastic for NN, but 
can be acute in other EFT applications in nuclear physics, 
e.g.\ where Coulomb barriers suppress low-energy cross sections so that accurate 
near-threshold data may be limited.

\item Distinguish between uncertainties stemming from  infrared (long-distance) and ultraviolet (short-distance) physics.
These two types of theoretical uncertainty will manifest themselves in different observables and in different kinematic domains. 
For example, errors in the NN interaction
associated with a long-distance parameter, such as
the pion-nucleon ($\pi$N) coupling constant, should be treated differently from uncertainties
in the parameterization of short-distance physics. Because the EFT explicitly separates short- and long-distance effects, it provides a straightforward way to do this, as long as 
separate errors for infrared and ultraviolet parameters are assigned. 

\item Facilitate tests of whether the EFT is working as advertised, i.e.\ whether the predictions of the EFT are improving as defined by the $Q$-expansion, and whether errors have the momentum dependence predicted by the naive expansion parameter formed by the ratio of momentum to the EFT breakdown scale. 
\end{enumerate}

\noindent
Here we lay out, in broad terms, a recipe
for uncertainty quantification and EFT validation that has these features. 
A number of the ideas we collect here have been previously considered or implemented 
in part (see, for example, Ref.~\cite{Griesshammer:2008aaa}). 
Yet the detailed aspects of this program remain to be worked out. Thus we also emphasize the questions that 
are---or should be---the subject of continuing investigation. 

First, in Sect.~\ref{sec:ingredients} we describe in more detail the ingredients
our recipe requires.
In Sect.~\ref{sec:Tools} we introduce the tools we will apply to these 
ingredients. 
The first of these is Bayesian 
methodology, which we advocate for uncertainty quantification 
because it allows us to include theoretical expectations
(such as LEC naturalness) in the analysis as prior 
information~\cite{Schindler:2008fh}, and, 
more generally, to treat consistently the various sources of error.
We also discuss error plots~\cite{Lepage:1997cs,Griesshammer:2004pe} as a 
prototypical analysis method to test whether the EFT is really providing a systematic expansion
in powers of $Q$ or not.
In Sect.~\ref{sec:procedures} we give a taste of how the recipe is
carried out 
by applying the tools to a toy problem from Ref.~\cite{Schindler:2008fh}.
We finish with a summary and outlook.

\section{Ingredients} \label{sec:ingredients}

The basic ingredients in our recipe are:
(\textit{i}) the Hamiltonian and associated operators;
(\textit{ii}) the method to solve for the observables of interest;
and (\textit{iii}) the LECs.
A similar list would exist for a non-EFT approach
to nuclear physics.  But in an EFT approach each ingredient is associated with special considerations regarding
uncertainty quantification.

 \subsection{The Hamiltonian} \label{sec:Hamiltonian}

Chiral perturbation theory ($\chiPT$) encodes the consequences of QCD at 
energies of order the pion mass. In particular it can be used to compute the 
interaction of single nucleons and pions for momenta well below the 
chiral-symmetry-breaking scale, $\LamchiSB$. 
A recent combined analysis of the 
level shift and width in pionic hydrogen and the level shift in pionic deuterium
using $\chiPT$ offers a good example of EFT uncertainty
quantification. 
These experimental observables were related to the isoscalar and isovector 
$\pi$N scattering lengths using a $\chi$PT Hamiltonian for $\pi$N interactions 
that had been systematically expanded in powers of $m_\pi/\LamchiSB$. 
The combined $1\sigma$ error ellipse yielded~\cite{Baru:2010xn}:
\begin{equation}
   \tilde{a}^+ = (1.9 \pm 0.8) \times 10^{-3} m_\pi^{-1} 
    \;, \quad
    a^- = (86.1 \pm 0.9) \times 10^{-3} m_\pi^{-1}
    \;.
\end{equation}
Ultimately this analysis provided the first definitive (2$\sigma$) evidence 
for a positive isoscalar $\pi$N scattering length.  
For both the $\pi^-$D and $\pi^-$H level shift a careful analysis of the 
impact of higher-order (in $m_\pi/\LamchiSB$) terms in the $\pi$N Hamiltonian 
was a crucial aspect of the work, since for both observables the experimental 
error is much smaller than this theoretical uncertainty. 

While $\chi$PT yields a purely perturbative expansion in powers of 
$(p,m_\pi)/\LamchiSB$ for the $\pi$N scattering amplitude, nuclei are bound 
states, and so cannot be generated from such an expansion. 
In the early 1990s Weinberg pointed out that the infrared enhancement associated
with multi-nucleon intermediate states meant that the $\chi$PT expansion could 
not be applied directly to the scattering amplitude in systems with 
more than one nucleon~\cite{Weinberg:1990rz}. He
argued that the $\chi$PT Lagrangian and counting rules should instead be used 
to compute an NN (or NNN or \ldots) Hamiltonian up to some fixed order, $n$, 
in $\chi$PT. Such an expansion can then be examined for convergence with $n$.
The $\chi$PT potential $V$ was computed to $O(Q^3)$ in 
Refs.~\cite{Ordonez:1995rz,Epelbaum:1999dj,Entem:2001cg}, 
and to $O(Q^4)$ in Refs.~\cite{Entem:2003ft,Epelbaum:2004fk}. 
Consistent three-nucleon forces have been derived and implemented in such an approach~\cite{VanKolck:1994yi,Epelbaum:2002vt}.

However, while there is a $\chi$PT expansion for $V$, the resulting nuclear binding energies
(and other observables) contain effects to all orders in the chiral expansion: there is
no obvious perturbative expansion for them. In practice, chiral EFT for few-nucleon systems
is often implemented as described in the previous paragraph, but with the Hamiltonian acting
only on a limited space: in 
momentum space a cutoff $\Lambda$ in the range $450 < \Lambda < 800$ MeV must 
be imposed~\cite{Marji:2013uia}.
From now on when we use the term chiral EFT in the context
of few-nucleon systems we mean calculations that are carried out in this way.

In such a calculation there are two different sources of uncertainty: when the cutoff $\Lambda$ is $\leq \breakdown$ it will remove physics that could, in principle, be described in the EFT, replacing it with a model that follows from the specific form of the regulator employed. In an $n$th-order EFT Hamiltonian the resulting errors in the short-distance physics should affect observables at worst as $(p/\Lambda)^{n+1}$. If these ``regulator artifacts" manifest themselves more strongly then the calculation is not properly renormalized. If it is, 
then taking $\Lambda \gg \breakdown$ allows for minimization of these artifacts.
Even then, it will remain the case that the EFT Hamiltonian has been truncated in the $Q$-expansion, and so the errors in the calculation will scale as $(p/\breakdown)^{n+1}$. 
Both $p/\Lambda$ and $p/\breakdown$ errors arising from the chiral EFT Hamiltonian must be considered when formulating error estimates, and in 
Sec.~\ref{sec:Lepage} we show how to disentangle them. But, when chiral EFT is   implemented only with $\Lambda \sim \Lambda_{\chi {\rm SB}}$, it is difficult to assess whether the EFT is behaving as desired with regard to its {\it a priori} error estimates, since $p/\Lambda$ errors cannot be cleanly separated from $p/\breakdown$ errors. For a review of works that remove this limitation, by computing the NN system in chiral EFT for cutoffs both of order and much greater than $\Lambda_b$,
see Ref.~\cite{Phillips:2013fia}.

Regardless, chiral EFT Hamiltonians will continue to be a  workhorse in 
nuclear-structure calculations  because they are 
(relatively) soft and provide a  rich set of operator structures 
for NN and 3N forces.
And we anticipate that their future incarnations will be improved via the inclusion of additional operators that help to remove $p/\Lambda$ errors,
and the inclusion of the $\Delta(1232)$ as an explicit degree of freedom---the latter ensuring that the pattern of EFT convergence is not deleteriously affected by this intermediate-energy excitation of the nucleon.

 \subsection{Methods for calculating} \label{sec:Methods}

The second ingredient in our recipe is the many-body approach or method (e.g., 
no-core shell model or coupled cluster)
that is used to calculate observables in nuclei.

Because the chiral EFT Hamiltonians used in practice effectively have lower 
cutoffs than high-precision phenomenological potentials, they are softer and hence generally more amenable to these methods. 
Further softening of potentials for nuclear structure and reaction calculations 
is often done using renormalization-group  
(RG) evolution~\cite{Bogner:2009bt}.  
EFT Hamiltonians are well-suited to this procedure, since RG transformations are 
an intrinsic part  of any EFT.
The similarity renormalization group (SRG), which applies 
unitary transformations to the Hamiltonian to change the effective 
resolution, is particularly popular because the evolution of three-body 
forces is technically straightforward~\cite{Furnstahl:2013oba}. 
In principle the unitary evolution under the SRG preserves all matrix 
elements of an initial EFT Hamiltonian, so $p/\Lambda_b$ truncation errors
do not increase even though the resolution scale is taken well below
$\Lambda_b$ (in contrast to an EFT with cutoff $\Lambda \ll \Lambda_b$). 
In practice this evolution induces new many-body forces.
Four- and higher-body forces are typically neglected at present, thus 
producing a new source of error for many-body observables.
This error and associated errors in the many-body method (which also
vary with resolution) can be assessed in part by tracking the dependence
of calculated observables on the resolution scale.  However, a more complete
method of assessing the uncertainty from omitted induced many-body forces
is desirable.

The evolution of Hamiltonians to softer, and therefore numerically more convergent, forms for 
many-body calculations raises the question of whether this advantage can be 
realized directly by designing an EFT potential specifically for the many-body 
method at hand. Stetcu and collaborators have developed the no-core shell model 
within an EFT framework~\cite{Stetcu:2006ey}. In their work the truncated 
harmonic-oscillator basis used to expand the many-body wave function also serves
as the UV regulator (it is an IR regulator as well). The EFT Hamiltonian is tuned 
to reproduce bound-state and scattering data exactly, irrespective of the 
model-space truncation. In this way the EFT errors and associated uncertainties 
in LECs become intertwined with the uncertainty of the many-body method. But the
calculation need not be any less accurate than if those errors were independent.
Similarly, in nuclear lattice simulations the coupling constants are functions 
of the lattice parameters~\cite{Lee:2008fa}; see the article by Epelbaum and Lee 
on uncertainty quantification in such a scheme elsewhere in this volume.

In contrast, as currently implemented for nuclear structure, most {\it ab initio} many-body methods do not distinguish between EFT and non-EFT Hamiltonians. Aspects of the Hamiltonian uncertainties  will depend on the EFT cutoff and the choice of regulator. This can, in turn, influence the convergence properties of the many-body calculation. Beyond that though, the EFT errors do not interact in any special way with the chosen many-body method, which then needs to have uncertainties that are either quantifiable or known to be smaller than other sources of theory error. The uncertainties associated with several of these methods are discussed in other contributions to this volume.

 \subsection{Low-energy constants} \label{sec:LECs}
 
Every EFT Hamiltonian contains undetermined parameters that summarize 
the low-energy effects of the degrees of freedom not explicitly included in 
the theory. The NN Hamiltonians used in chiral EFTs contain $\pi$N LECs, 
which encode short-distance dynamics in the $\pi$N system. This, in turn, 
controls NN dynamics at distances $\sim 1/m_\pi$. The most important of these 
is the combination 
that gives the 
$\pi$N coupling constant at the physical pion mass. As was demonstrated 20 
years ago, the value of this coupling constant can be precisely determined by 
analyzing the energy dependence of NN data~\cite{Stoks:1992ja}. One-pion 
exchange generates rapid energy dependence in the NN data, which can be 
reliably separated from the slower energy dependence produced by physics at 
$r \sim 1/\LamchiSB$.  In the formulation of multi-nucleon chiral EFT 
originally proposed by Weinberg~\cite{Weinberg:1990rz} and implemented in Refs.~\cite{Ordonez:1995rz,Epelbaum:1999dj,Entem:2001cg,Entem:2003ft,Epelbaum:2004fk} there
  are 9 NN LECs in the $O(Q^3)$ $H$ 
and 24 such LECs in the $O(Q^4)$ $H$; these encode the NN short-distance dynamics. 

These LECs typically
have been fit to laboratory NN phase shifts for cutoffs $\Lambda$ in the 
range 500--800 MeV (450--600 MeV at $O(Q^4)$).\footnote{New fits are in progress
that regulate the long-distance physics with a local, coordinate-space
regulator, which reduces regulator effects on long-distance pion exchange.} 
The resulting LECs are ``natural", in the sense that they are 
of the size expected from arguments regarding their scaling with the pion decay 
constant $f_\pi$ and $\LamchiSB$~\cite{Georgi:1992dw}.  Thus, in this $\Lambda$ range, the size of the LECs obtained in fits to NN data 
is in accord with the underlying hadronic scales. 
Resonance saturation has also been used to account for the broad pattern of 
these LECs~\cite{Epelbaum:2001fm}---although it must be applied
with care in non-perturbative contexts. {\it A posteriori}, at least, 
naturalness is consistent with the value of the NN LECs that enter these 
potentials.    

In the longer term we can anticipate extracting LECs from lattice QCD simulations of the NN system.  But lattice QCD cannot ``measure" LECs directly: it can only compute S-matrix elements. So LECs must always be extracted from data, regardless of whether that data comes from the laboratory or from numerical solutions of QCD.  
In the NN case there has been much discussion as to the  energy range over which data should be fit. In principle the EFT should be
fit to the most infrared data possible~\cite{Lepage:1997cs}, as the
EFT errors grow with $p$.
The (non-EFT) Nijmegen analysis referred to above fit NN data up to 350 MeV of lab energy, whereas Ref.~\cite{Epelbaum:1999dj} worked to $O(Q^3)$ in chiral EFT and determined LECs using data at $E_{\rm lab} \leq 100$ MeV~\footnote{There is also the complementary problem of ensuring that the database being used to extract the LECs is statistically consistent. We do not discuss this further here, but refer to Ref.~\cite{Perez:2014yla} for a presentation of tests that permit the removal of non-statistical components of the NN database.}
. More recently, the $O(Q^3)$ computation of Ref.~\cite{Ekstrom:2013kea} fit NN data to 125 MeV, accounting for the growth in EFT uncertainty over this range by adding a $p$-dependent piece to the weights in the objective function (cf. Eq.~(\ref{eq:chisqwithEFTerrors}) below). Different choices underlying the extraction of NN LECs from data affect all subsequent predictions of the theory.  Ideally, there would be no arbitrary decision regarding the energy range of the data used to fix the LECs. Instead, by accounting for the theoretical uncertainty in a rigorous way, we would automatically optimize the trade-off between more (and often better) data at higher energies, and the decrease in the accuracy of the EFT Hamiltonian as $p$ gets larger. 

\begin{figure}[tbh-]
 \begin{center}
 \includegraphics[width=0.9\columnwidth]{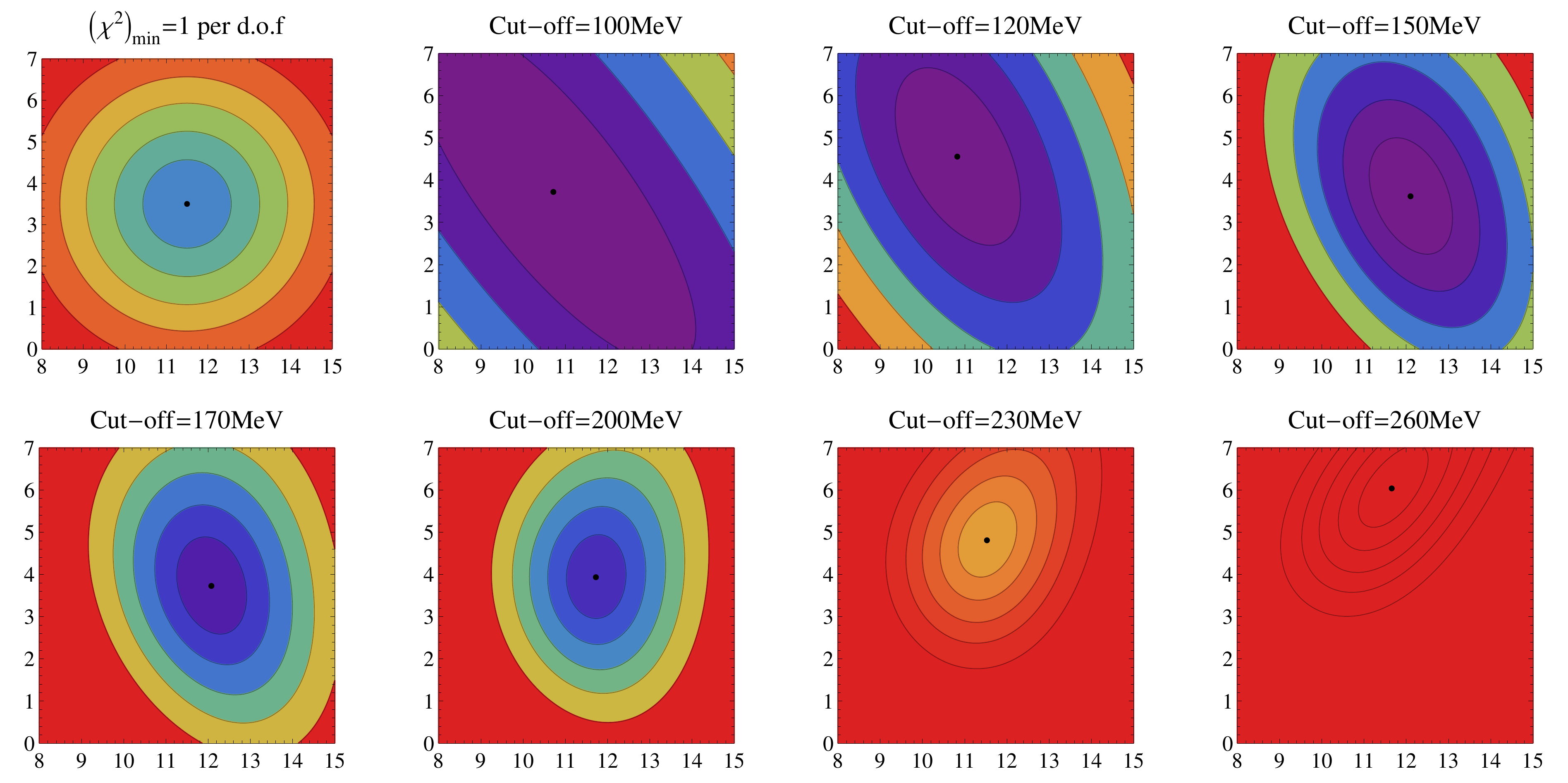}
\caption{Likelihood contours in the $\alpha_{E1}^{\rm(p)}-\beta_{M1}^{\rm (p)}$ plane from fits to $\gamma$p differential-cross-section data up to seven different maximum values of  energy and momentum transfer (``cut-offs"), as indicated in the captions. The first panel shows the color scale, with the minimum artificially set to  $\chi^2/{\rm d.o.f.}=1$. In each panel the contours correspond to confidence levels of 68\%, 95\%, and so on.
The units of the $\alpha_{E1}^{\rm (p)}$ $(x)$ and $\beta_{M1}^{\rm (p)}$ $(y)$
axes are $10^{-4}$ fm$^3$. For details of the data base and theory see Refs.~\cite{Beane:2004ra,McGovern:2012ew}. 
Figure courtesy J.~A.~McGovern. 
}
 \label{fig:ellipses}
 \end{center}
\end{figure}

Up until now the best that could be done in this regard was to fit data to different maximum energies, and ensure that the fit was stable for a range of choices. Such a test was done for a $\chi$PT analysis of proton Compton scattering data~\cite{Beane:2004ra,McGovern:2012ew}. There the fit parameters are the proton electric and magnetic dipole polarizabilities, $\alpha_{E1}^{\rm(p)}$ and $\beta_{M1}^{\rm (p)}$, each of which contains one $\chi$PT LEC. These were fixed by fitting data up to different energy cutoffs. Fig.~\ref{fig:ellipses} shows that the  likelihood contours in the $\alpha_{E1}^{\rm(p)}-\beta_{M1}^{\rm (p)}$ plane become tighter as the maximum energy of the fit is increased. However, the fit becomes both unstable and worse once $E_{max} > 200$ MeV, because this EFT is unable to describe data at energies and momentum transfers where the $\Delta(1232)$ plays a marked role.

There are also LECs in the three-nucleon force. Some of these can only be extracted in fits to properties of systems beyond A=2. Therefore, it will never be the case 
that two-nucleon data alone can determine the nuclear force.
But it may also be the case that certain NN (or even $\pi$N) LECs are better determined from calculations in the three-, four-, \ldots nucleon system. 
And indeed, NN data has already been used to constrain the $\pi$N LECs $c_3$ 
and $c_4$ in chiral EFT in Refs.~\cite{Entem:2002sf,Rentmeester:2003mf}. 
However, these two sets of authors obtained values for $c_3$ and $c_4$ which differed by more than the quoted errors, which implies that the theoretical uncertainties of at least one of these chiral EFT calculations are underestimated. This is a cautionary tale for future efforts to determine an LEC in systems larger than that in which it first appears. Such an approach may make excellent sense if precise data are available which are particularly sensitive to the LEC in question---as is the case for $c_3$ and $c_4$ in the NN system. But it is mandatory that the systematics of the calculation in higher-body systems be understood, or otherwise different extractions of the same LEC
will be inconsistent---throwing into question the ability of chiral EFT to provide a systematic nuclear force.

Another issue complicates the choice of observables to constrain LECs. For many nuclear reactions, certain reaction
observables will be poorly predicted if the separation energy is not accurately reproduced.  In the case of the NN system the convergence of low-energy deuteron observables is markedly improved if we fit the EFT LECs to the deuteron binding energy and asymptotic S-state normalization rather than to, say, the ${}^3$S$_1$ scattering length and effective range~\cite{Phillips:1999hh}. However, it should be noted that these two fit choices are equivalent up to higher-order terms, and so results obtained with them
should be consistent within statistical uncertainties and higher-order effects.

\begin{figure}[tbh-]
 \begin{center}
 \includegraphics[width=0.65\columnwidth]{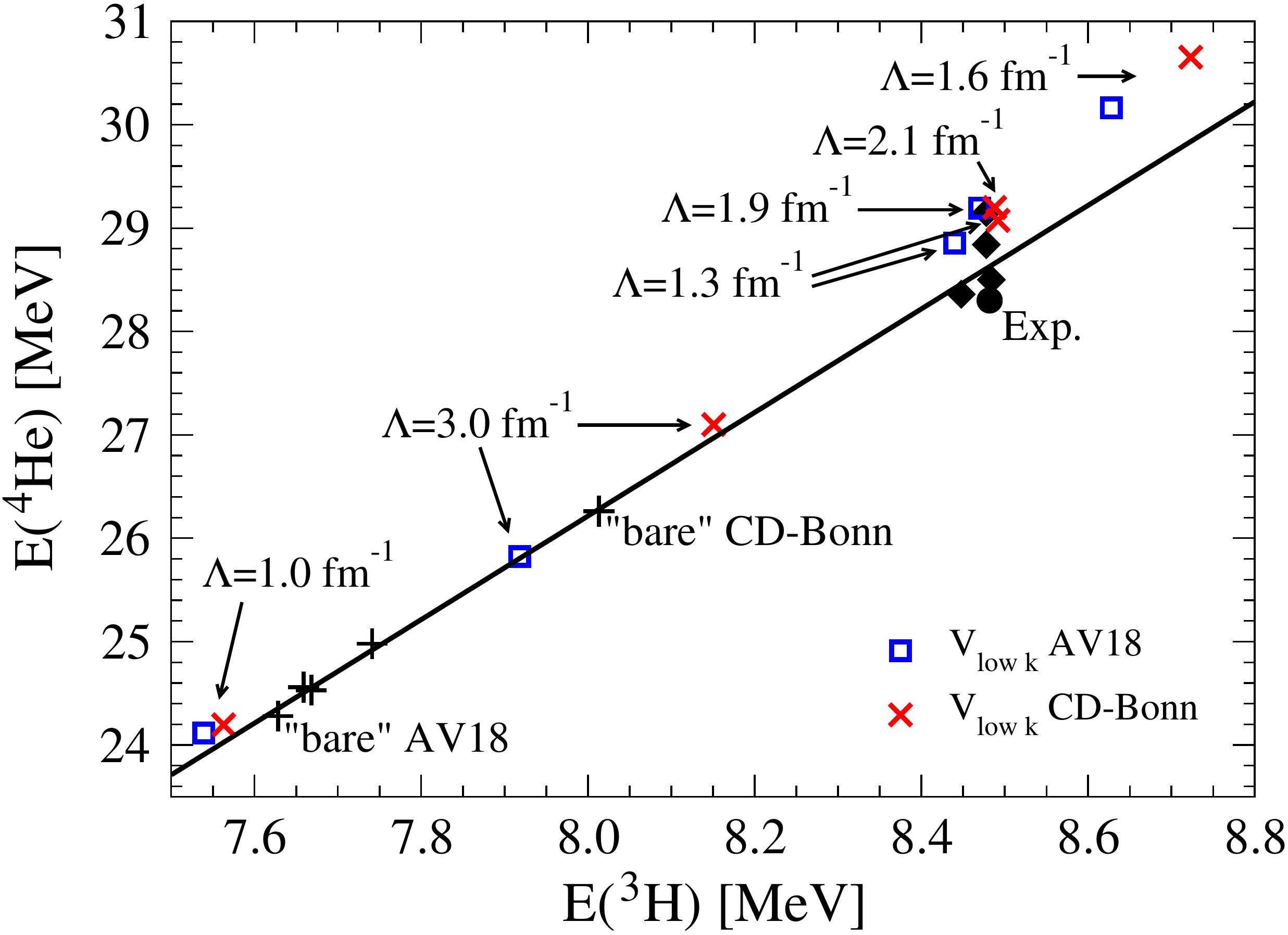}
\caption{The Tjon line correlation between $^4$He and $^3$H binding energies
        for several phenomenological potentials (plus signs) and 
        for potentials evolved by renormalization group methods from
        two of them (squares and crosses)~\cite{Nogga:2004ab}.  
        The diamonds include adjusted three-body forces. The
        experimental values are shown as a circle.}
 \label{fig:correlations}
 \end{center}
\end{figure}

In this sense tuning an incomplete EFT calculation (e.g.\ one with NN forces only) to reproduce the separation energy of a nucleus very precisely only yields accurate results for a low-energy reaction observable insofar as that observable is correlated with the separation energy. Such a calculation has arranged for a particular class of neglected EFT effects to cancel in the separation energy, but there is no guarantee such a cancellation will persist in other, non-correlated, nuclear observables.  Moreover, one particular member of a family of Hamiltonians has been singled out: the full family of Hamiltonians that exists at this order is not being sampled. Thus, unless there is prior evidence for such a correlation between the separation energy and the quantity of interest this kind of fine tuning may be dangerous. 

It can also be dangerous because these types of correlations could lead to degeneracy in the observables chosen for LEC extraction, thereby causing instabilities that drive the LECs to unnatural values. This may be happening with some contemporary fits of $c_D$ and $c_E$, e.g.\ if these are fit to the triton binding energy and the doublet neutron-deuteron scattering length, which are known to be correlated via the ``Phillips line"~\cite{Gazit:2014}. For similar reasons it would seem unwise to include both the triton and alpha-particle binding energies in a fit, since these two observables can be expected  to produce degenerate information in this sense (see Fig.~\ref{fig:correlations}). 
On the other hand, such correlations can often be recognized {\it a priori} using very-low-energy EFTs (e.g.\ Halo EFT), providing guidance to the chiral EFT fit.

\section{Tools} \label{sec:Tools}

In this section we give brief descriptions of some tools we will use in our recipe for EFT uncertainty quantification.  
Some of the details are illustrated by the example in Sec.~\ref{sec:procedures}.

\subsection{Bayesian methods} \label{sec:Bayesian}

\newcommand{\avec}{\mathbf{a}}

Our recipe for uncertainty quantification must incorporate the special EFT ingredients
we have described.
Following Ref.~\cite{Schindler:2008fh},
we propose a Bayesian framework to encode the expectations inherent in an EFT
as prior information.

The literature on Bayesian methods and applications is diverse and growing;
we recommend Ref.~\cite{Sivia:06} for a gentle but thorough introduction.
Here we confine ourselves to those elements relevant for EFT applications,
and specialize the notation to this end.
Thus, we denote a chosen set of observables (data) by $D$, the 
information on experimental errors and on the EFT
(specified by an order, renormalization and regularization schemes, 
low-energy degrees of freedom and symmetries, etc.) by $I$, and the vector of LECs 
for the EFT by $\avec$.  

Bayes theorem itself is a straightforward consequence of the laws of 
conditional probability.
In the present case it states that
\beq 
  \pr(\avec|D,I) 
      = \frac{\pr(D|\avec,I) \, \pr(\avec|I)}{\pr(D|I)}  \;.
  \label{eq:bayesthm}
\eeq
The conventional terminology for the probabilities in Eq.~\eqref{eq:bayesthm}
is~\cite{Sivia:06}:
\begin{itemize}  
 \item$\pr(\avec|D,I)$ is the \emph{posterior} probability distribution
 function (pdf), which represents the 
 probability of the parameters given the data and the model, with all 
 information taken into account;
 \item $\pr(D|\avec,I)$ is the \emph{likelihood} pdf. For normally
 distributed errors on independent measurements, it is given by 
 $e^{-\chi^2/2}$ with $\chi^2$ the usual chi-squared function;
 \item $\pr(\avec|I)$ is the \emph{prior} pdf, which represents 
 knowledge we have about the parameters before the measurement of
 the data; 
 \item ${\pr(D|I)}$ is the \emph{evidence}.
\end{itemize}
The posterior  lets us find, given the data we have measured,
the most probable values of the parameters and to predict
values of observables with confidence intervals.

Bayes theorem thus relates the probability of measuring certain
data, given a specified EFT and a given set of LECs, to the converse: 
the probability that the parameters are correct
given the data and the EFT. 
It is in the interpretation of probability densities such as $\pr(\avec|D,I)$ and $\pr(\avec|I)$
that controversy arises. In particular, a probability for parameters to fall in certain ranges (``credibility interval") makes little sense in a frequentist interpretation. 
Instead, these intervals are understood to represent the ``degree of belief'' as to whether the true values of the parameters fall in that interval or not. These notions
can be quantified using concepts such as the ``coherent bet"~\cite{deFinetti:1974}. 
The use of a prior pdf in this process may appear subjective, but $\pr(\avec|I)$ could simply summarize results from previous experiments. 
Furthermore,
the maximum-entropy principle can be used to develop priors
that implement certain pieces of ``testable information" in a way that minimizes any additional information imparted to the prior~\cite{Jaynes:1957}.
Regardless of how $\pr(\avec|I)$ is obtained,
Eq.~\eqref{eq:bayesthm} is a mathematical statement
regarding how our knowledge of $\avec$ is updated when new data appears.  Thus if the (new) data are numerous, and/or of high quality, they will 
overwhelm any reasonable
prior, rendering it insignificant in the computation 
of the posterior pdf. 

The toy problem in Sec.~\ref{sec:procedures} provides an instructive contrast between
conventional and Bayesian approaches.  It focuses on \emph{parameter estimation}; the goal is the posterior distribution
for $\avec$,
which is obtained by specifying the prior and the likelihood (for parameter estimation the evidence is just a
normalization).  If we assume Gaussian errors for the data and
non-informative (e.g., constant) priors, the determination of the most likely
value reduces to the
familiar $\chi^2$ minimization.
Some types of theoretical error can be included by adding an
appropriate ``penalty'' to the likelihood~\cite{Dobaczewski:2014jga}.

We follow Ref.~\cite{Schindler:2008fh} and argue that a well-chosen prior can add useful information to the fit when data is limited. In particular, imposing 
naturalness on EFT fits (i.e. the principle that when rescaled as dimensionless variables, the LECs should be of order one) can aid both the stability of the fit and
the quantification of uncertainties.
In Ref.~\cite{Schindler:2008fh} a maximum-entropy prior was used to encode the 
theoretical expectation of natural-sized LECs. That work also emphasized that 
marginalization~\cite{Sivia:06} provides a straightforward way to account for higher-order EFT contributions;
 their coefficients can be treated as ``nuisance parameters". For example, if we only wish to determine $p+1$ LECs, $a_0$, $a_1$, \ldots, $a_p$,
we compute:
\begin{equation}
\pr(a_0,a_1,\ldots,a_p|D,I)= {\cal N} \int \pr(D|\avec,I) \, \pr(\avec|I)\, d{\bf a}_{\rm marg}
\; ,
\label{eq:marg}
\end{equation}
where the vector ${\bf a}_{\rm marg}$ contains the parameters $a_{p+1}, \ldots, a_M$, and ${\cal N}$ is a normalization. $M$  is the order of the fit, and
marginalization over $M$ can also be performed. Marginalization can also be used to avoid
rigid specification of the naturalness prior. 

Besides parameter estimation,
Bayesian methods permit a quantitative approach to {\it model selection}: the
choice between two different 
models with different parameter sets~\cite{Sivia:06}. In the case that one set of parameters is a subset of the other this provides
a mathematical implementation of Occam's razor: new parameters should be added to the model only
if they markedly improve the agreement with data. It can be used, e.g., to decide whether a linear or quadratic function is a better model for a set of data~\cite{Sivia:06,Jeffreys:1939}.
In Bayesian model selection, the key quantity is the evidence ratio for the
two different models $M_1$ and $M_2$, given the same data $D$, and common assumptions $I$~\cite{Jeffreys:1939}:
\begin{equation}
  \frac{\pr(M_1|D,I)}{\pr(M_2|D,I)} =
  \frac{\pr(D|M_1,I)\,\pr(M_1,I)}{\pr(D|M_2,I)\,\pr(M_2,I)} 
    \;.
    \label{eq:evidence_ratio}
\end{equation}
If both models are {\it a priori} equally likely given the assumptions $I$ then the model-evidence 
ratio reduces to the Bayes ratio, for which we must integrate (marginalize) over the full range of the (different)
sets of parameters $\avec_1$ and $\avec_2$:
\begin{equation}
  \frac{\pr(D|M_1,I)}{\pr(D|M_2,I)} =
    \frac{\int \pr(D|\avec_1,M_1,I)\,\pr(\avec_1|M_1,I)\,d\avec_1}
         {\int \pr(D|\avec_2,M_2,I)\,\pr(\avec_2|M_2,I)\,d\avec_2}
    \;.
    \label{eq:Bayes_ratio}
\end{equation}

Examples from Ref.~\cite{Sivia:06} suggest that model-selection criteria could be used to 
decide which EFT parameters actually improve the fit to data, which could, in turn,
lead to concrete tests of different choices regarding the pertinent low-energy
scales in the EFT, i.e. different EFT power countings. In particular, the EFT 
should be able to distinguish improvements in the fit that arise from correctly identifying short-
and long-distance physics from those which occur simply because there are more 
parameters as one moves to higher order.  
While we do not yet have a concrete formulation of such tests, the 
evidence ratio suggests quantitative criteria can be developed, and
discussions surrounding these issues need not be restricted to 
vague statements such as ``it looks like it is working". 

Our advocacy of a Bayesian framework for EFT uncertainty
quantification is supported by the widespread application of
Bayesian approaches in physics. For example: the interpretation of
dark-matter searches~\cite{Arina:2011si,Pato:2012fw}; structure determination in condensed-matter physics~\cite{Rieping:2005};
and constrained curve-fitting in lattice QCD~\cite{Lepage:2001ym}.
Two examples deserve special mention here. First, 
questions about
whether supersymmetry is a ``natural'' approach to the hierarchy problem
have led several authors to impose Bayesian priors in 
favor of natural values of the constants in supersymmetric models~(see Refs.~\cite{Cabrera:2008tj,Ghilencea:2012qk} for two examples).
By combining those priors with data (primarily from the LHC), they hope to
quantitatively determine the extent to which the posterior pdf constrains 
the parameter space for supersymmetric models. And in Ref.~\cite{Balazs:2013qva} the Bayes ratio
is used to compute the evidence for the constrained minimal supersymmetric standard model; in that work naturalness
informs one of the priors used to calculate the ratio. 
Despite the similar nomenclature, naturalness in nuclear EFTs plays a very different role.
Our issue is to decide which data are most efficacious for extracting the 
LECs of the theory because, as described above, 
as the energy range over which the fit is done 
increases there is a trade-off between more data and a theory in which 
omitted higher-order terms play an increasingly prominent role. 

A more relevant analogy for the 
theoretical uncertainties in nuclear EFTs is the need 
to estimate theoretical
uncertainties in processes computed with parton distribution functions (PDFs) 
at a given order due to missing higher-order perturbative QCD (pQCD) 
corrections. Bayesian approaches have recently been applied
to estimate these uncertainties~\cite{Cacciari:2011ze,Forte:2013mda}.
This is particularly important in cases where the standard technique of varying the 
scale parameter in the pQCD calculation underestimates the size of 
higher-order terms, e.g.\ Higgs production by gluon-gluon fusion. Information on the first few pQCD coefficients can be used to formulate
a Bayesian prior from which the theoretical uncertainty can be derived using 
straightforward mathematics and Eq.~(\ref{eq:bayesthm}). 
The case of nuclear physics is more complicated, though, as
it is intrinsically non-perturbative. The coefficients (LECs)
must be extracted from data; they cannot be calculated {\it a priori}.

\subsection{Analysis of systematic residuals} \label{sec:Lepage}

An EFT does not just provide a good fit to data. The residuals obtained when a finite-order EFT is compared to data {\it should} display a systematic trend: they should not 
be distributed statistically around zero, but instead should behave in a manner consistent with the dominant omitted terms. 
(This is in addition to the statistical residuals arising from experimental
measurements.)
Moreover, as the order of the EFT calculation is 
increased, residuals should decrease in accord with the power counting. 
Thus, if it is working as advertised, the EFT's predictions
should 
\begin{itemize}
  \I at a given  order, degrade with increasing $p$ as expected on the basis of the  power counting;
  \I improve order-by-order following the assumed power counting.
  \end{itemize}
Moreover, it should be possible to infer the breakdown scale of the EFT from both of these trends.

A tool that can be used to analyze whether this is indeed the case
is the error plot, such as those used in
numerical analysis (e.g. see Fig.~6.3 in Ref.~\cite{landau2011survey}).
These were introduced to analyze EFT behavior  in
Ref.~\cite{Lepage:1997cs} and are commonly called ``Lepage plots''.
Error plots can help to disentangle and discern the size of the
 two types of errors in an $n$th-order EFT calculation: (a) ``regulator artifacts" should grow at worst like
 $(p/\Lambda)^{n+1}$ for $\Lambda < \Lambda_b$---as long as the calculation is properly renormalized;
 (b) the truncation of the EFT Hamiltonian at order $n$ means $H$ itself is only accurate up to terms 
$\sim (p/\breakdown)^{n+1}$.

\begin{figure}[tbh-]
 \begin{center}
  \includegraphics[width=0.49\columnwidth]{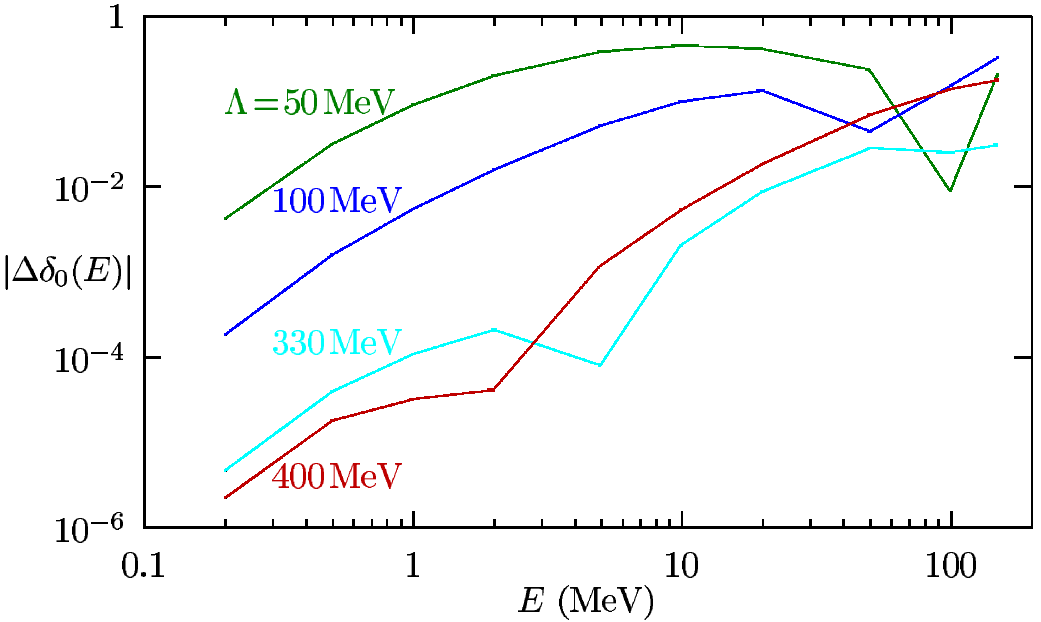}~~~~%
  \includegraphics[width=0.49\columnwidth]{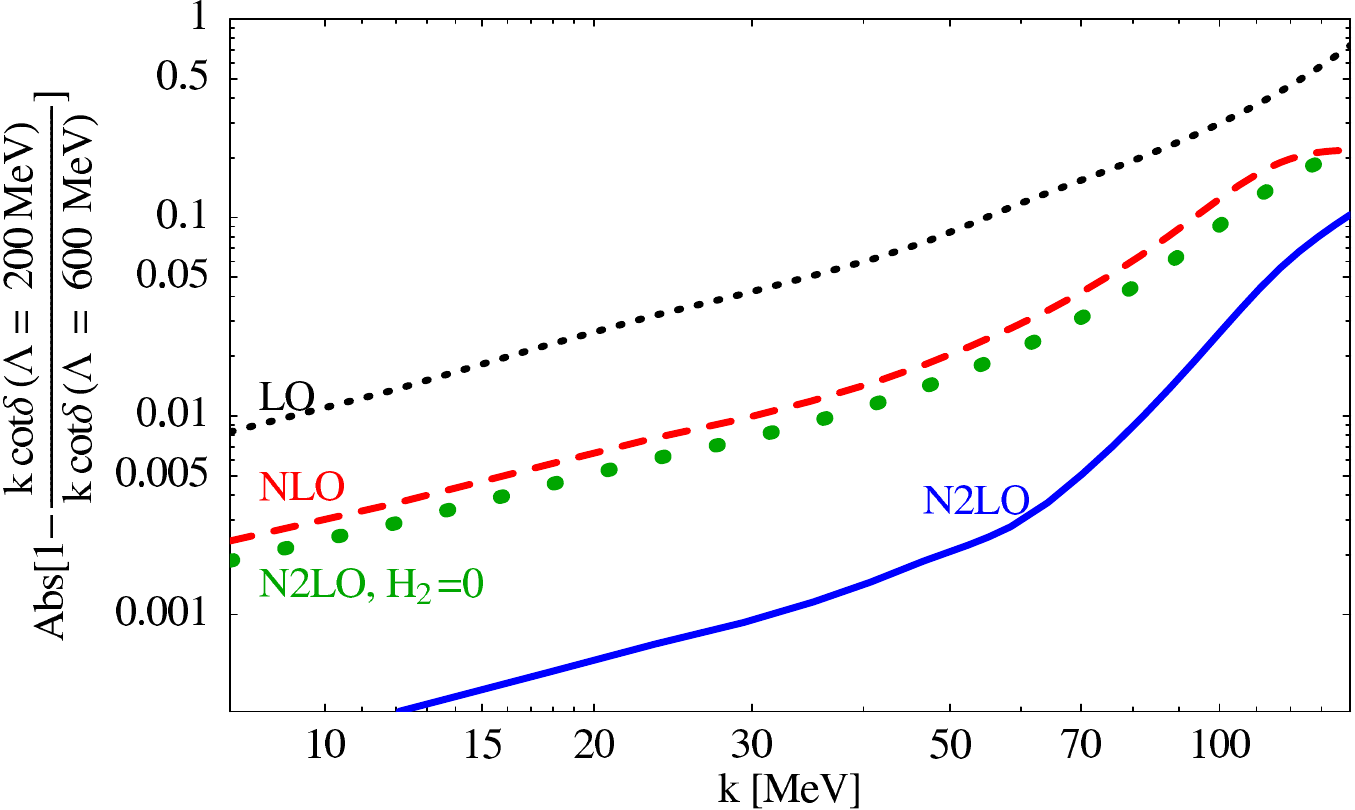}
 \caption{Left: error plot from Ref.~\cite{Lepage:1997cs} showing the residuals 
 of the neutron-proton 
${}^1$S$_0$ phase shift as a function of laboratory energy for an incomplete NLO 
calculation in chiral EFT using different values of the cutoff $\Lambda$.
 Right: error plot for neutron-deuteron scattering in the ${}^2$S$_{1/2}$ channel: here the comparison
 is between pionless EFT at two different cutoffs, and is done for several different orders in the pionless EFT expansion~\cite{Griesshammer:2004pe}. 
 }
 \label{fig:lepage_plots}
 \end{center}
\end{figure}

Examples of plots which quantitatively elucidate these two types of error 
are given in Fig.~\ref{fig:lepage_plots}.
On the left, the absolute value of the residual of NN
phase shifts (theory minus experiment) is plotted as a function of energy 
for a set of EFT calculations that all 
include two LECs for the short-distance NN physics in the $^1$S$_0$
channel, but employ different values of the momentum cutoff $\Lambda$.
(The calculations also include one-pion exchange, but not two-pion exchange. 
In terms of the standard chiral EFT power counting they are
``incomplete NLO'' calculations.) 
There are two aspects of such plots: 1) the slope of the residual with
energy (or momentum) at different orders should reflect the truncation
error, and 2) residuals should decrease at a fixed energy with increasing 
$\Lambda$ 
until the intrinsic cutoff scale is reached. The dominant errors seen here, then, 
are due to cutoff artifacts. 
The residuals
do not go to 
zero once $\Lambda > \breakdown$, but truncation error becomes the dominant effect
there.
In principle, the 
breakdown scale $\breakdown$
could be approximately determined from either trend: when lines from different
orders intersect the order-by-order expansion is no longer converging, and
cutoff artifacts stop decreasing for cutoffs $\Lambda$ around the breakdown scale
of the theory.

However, data is frequently sparse and often noisy: trying to discern from Lepage plots whether power-law exponents are in accordance with
EFT expectations can be decidedly difficult. A different type of error plot which does {\it not} involve a comparison to 
data---and reflects a different philosophy to the plot in the left panel---is shown in the right panel of 
Fig.~\ref{fig:lepage_plots}.
Here the relative difference of EFT S-matrix elements, specifically 
$k\cot\delta(k)$,   at two different cutoffs is plotted versus 
momentum, $k$~\cite{Griesshammer:2004pe,Bedaque:2002yg}.  (The pertinent LECs were fit
to the same low-energy $nd$ data at each cutoff.)
The two cutoffs are deliberately both chosen  above the ``pionless EFT"
breakdown scale, i.e. $\Lambda > 140$ MeV. 
The difference $k\cot(\delta(k))|_{\Lambda=600~{\rm MeV}}-k\cot(\delta(k))|_{\Lambda=200~{\rm MeV}}$ then reflects the 
size of the $(k/\breakdown)^{n+1}$ error: the coefficient of that error gets set in an uncontrolled way
to two different numbers in the two calculations. 
The slope of this difference above 70 MeV then becomes larger order-by-order: fits show 
the omitted operators
at LO, NLO, and N$^2$LO behave as $(p/\breakdown)^{n+1}$, with $n$ as expected from pionless-theory power counting~\cite{Griesshammer:2004pe}~\footnote{Below 70\,MeV the error involves scales 
other than $k$, so the scaling is not as straightforward~\cite{Griesshammer:2004pe}.}.

The behavior seen in these plots should be reflected in the error bands of an EFT calculation.  At $n$th order in the EFT their width should grow with $p$ as $(p/\Lambda_b)^{n+1}$, as long as $\Lambda > \breakdown$. It follows the width should decrease with increasing order even at fixed $p$. 
 However, this behavior will only be seen if the calculation has been renormalized properly, ensuring that regulator artifacts really are negligible for $\Lambda > \breakdown$.
Unfortunately such renormalization has not been carried out for  the most sophisticated contemporary chiral EFT Hamiltonians; they are implemented only for $500 < \Lambda < 800$ MeV because regulator artifacts play a dominant and destructive role for larger cutoffs. Recent attempts to implement NN chiral EFT such that regulator artifacts are negligible for $\Lambda > \Lambda_b$, i.e. the theory is properly renormalized, are summarized in Ref.~\cite{Phillips:2013fia}.

We note that analysis using plots like those in Fig.~\ref{fig:lepage_plots} is not the only way to glean information from the EFT fit.
Standard statistical techniques can be helpful, e.g.\ the singular value decomposition can reveal correlations between EFT parameters,
as discussed elsewhere  in this volume and in Ref.~\cite{Dobaczewski:2014jga}.

\section{Sample tasting: a toy example of fitting procedures and error
          analysis} \label{sec:procedures}

In this section we adapt our recipe for EFT uncertainty quantification
to a pedagogical toy problem from Ref.~\cite{Schindler:2008fh} 
to give a taste of Bayesian procedures.
This problem mimics 
some basic features of determining LECs from fits to data, so that we
can contrast the use of priors in finding posterior distributions 
with more conventional least-squares fitting procedures.
By generalizing the problem we can explore features and tools
that are not yet established and tested for chiral EFT 
interactions~\cite{Furnstahl:2014aa}.

The original toy problem is to find the leading coefficients in the 
Taylor expansion of a contrived function~\cite{Schindler:2008fh}
(e.g.\ simulating a cross section):
\bea
   g(x) &=& \left( \frac12 + \tan(\frac{\pi}{2}x) \right)^2  \nonumber \\
        &\approx& \sum_{i=0}^{M} a_i x^i 
           \approx 0.25 + 1.57x + 2.47x^2 + 1.29x^3 + 4.06x^4 + \cdots
       \;,
       \label{eq:toy_eq}
\eea
from various sets of artificial data to which gaussian noise is added
to simulate experimental errors.
In particular, the data are generated by choosing a set of $N$
evenly spaced $x$ points up to a maximum value of $x=1/\pi$ or $x=2/\pi$ 
(for example);
given the $j^{\rm th}$ point $x_j$, the data point $d(x_j) \equiv d_j$ and error 
$\sigma_j$ are
\beq
   d_j = g(x_j) (1 + c\eta_j)  
    \quad \Longrightarrow \quad \sigma_j = c \, d_j
   \;,
   \label{eq:pseudodata}
\eeq
where $\eta_i$ is normally distributed about 0 with standard deviation 
$\sigma_\eta = 1$,
and $c$ is a specified relative error (chosen initially to be $5\%$).
From Eq.~\eqref{eq:toy_eq} it is suggestive that the coefficients $a_i$
are of order unity, which operationally means between 1/4 and 4,
in analogy to the expected naturalness of LECs in an EFT.
Here the
radius of convergence of the Taylor series is 1, which
corresponds to the breakdown scale of the theory.
This toy problem in parameter estimation
is helpful in understanding the difficulties that come with 
fitting data to a model that deteriorates as the expansion parameter increases. 
The problem is particularly simple because it removes the complication of the 
non-linearity between the observables and the LECs in a chiral EFT,
but this feature is easily relaxed. 

Bayes theorem in Eq.~\eqref{eq:bayesthm} applies directly with 
the coefficient vector $\avec = \{a_i\}$ ($i=0$ to $M$), the data 
$D = \{d_j\}$ ($j=1$ to $N$), and with $I$ including the naturalness of
the polynomial coefficients and knowledge about the experimental errors.
Let us first neglect the information about naturalness and omitted
higher-order terms, i.e. choose a constant distribution for the prior $\pr(\avec|I)$
The evidence $\pr(D|I)$ is independent of $\avec$, so it disappears 
into the normalization of the posterior and can be dropped without consequence.
Then Bayes theorem tells us that the probability of a particular
set of coefficients $\avec$, given the data $D$, is equal to the
probability that these data would be measured given that the model of 
a polynomial is exact and that $\avec$ are its coefficients.

With our knowledge that the data are independent, the principle
of maximum entropy gives the likelihood pdf as~\cite{Schindler:2008fh,Sivia:06}
\beq
  \pr(D|\avec,I) = \prod_{j=1}^{N} \frac{e^{-\chi^2/2}}{\sigma_j \sqrt{2 \pi}}
    \;,
    \mbox{\ \ with\ \ }
    \chi^2 = \sum_{k=1}^{N}
    \left( \frac{d_k - \sum_{i=0}^M a_i (x_k)^i}{\sigma_k} \right)^2
    \;.  
\eeq
Taking the maximum of the likelihood pdf as our best result
is then the same as minimizing $\chi^2$, and we recover a standard
least-squares procedure within the Bayesian formulation.
(Note that we are obtaining the values for the $a_i$'s that make the data
most likely.)
But, in practice, there are difficulties, as seen in the sample
results from Ref.~\cite{Schindler:2008fh}
in the left table of Table~\ref{tab:toy_results1}.
The results for the $a_i$'s become increasingly unstable as
$M$ rises, despite good $\chi^2$/dof values,
and there is no clear way to determine which fit is optimal
(see Refs.~\cite{Schindler:2008fh,Furnstahl:2014aa} for further discussion
and more examples).

\begin{table}[thb!]
    \renewcommand{\tabcolsep}{2pt}
 \begin{center}   
     \begin{tabular}{|c|c|c|c|c|}
     \hline
     M & $\chi^2$/dof & $a_0$  & $a_1$  & $a_2$ \\ \hline
    1  & 2.24 & 0.20$\pm$0.01  &  2.55$\pm$0.11 & \\ 
  2 &  1.64 & 0.25$\pm$0.02  &  1.57$\pm$0.40  &  3.33$\pm$1.31  \\ 
  3 &  1.85 & 0.27$\pm$0.04  &  0.95$\pm$1.10  & 8.16$\pm$8.05   \\
  4 &  1.96 & 0.33$\pm$0.07  &  $-$1.88$\pm$2.69  & 44.7$\pm$32.6 \\  
  5 & 1.39 & 0.57$\pm$0.13  &  $-$14.8$\pm$6.85  & 276$\pm$117 \\ \hline  
     \end{tabular}
  \hspace*{.3cm}
     \begin{tabular}{|c|c|c|c|c|}
     \hline
     M &  $a_0$  & $a_1$  & $a_2$ \\ \hline
    1  & 0.20$\pm$0.01  &  2.55$\pm$0.11 & \\ 
    2  & 0.25$\pm$0.02  &  1.63$\pm$0.39  &  3.15$\pm$1.27  \\ 
    3  & 0.25$\pm$0.02  &  1.65$\pm$0.45  &  2.98$\pm$2.32   \\
    4  & 0.25$\pm$0.02  &  1.64$\pm$0.46  &  2.98$\pm$2.39 \\  
    5  & 0.25$\pm$0.02  &  1.64$\pm$0.46  &  2.98$\pm$2.39 \\ \hline  
     \end{tabular}
 \end{center}

 \caption{Left: results for the leading coefficients
 of the Taylor expansion in Eq.~\eqref{eq:toy_eq} for
 $\chi^2$ minimizations that fit polynomials of order $M$. 
 (Note that the agreement for $M=2$ of $a_0$ and $a_1$ with
 the exact expansion is coincidental.)
 Right: results including a normal Bayesian prior with
 $R=5$. The synthetic data used in the fitting procedures was generated 
 with $c=0.05$ and $x_{max} = 1/\pi$: 
 see dataset ``D$1$" in \cite{Schindler:2008fh}.
 }

 \label{tab:toy_results1}
\end{table}

The first step to overcome this problem is to evaluate the 
posterior distribution
with a prior that encodes our knowledge (or expectation) that each of
the $a_i$'s is of order unity.  To do so we need to adopt a functional
form for the prior; we follow Ref.~\cite{Schindler:2008fh} and assume
 a normal distribution
for each coefficient with mean zero and width $R$. 
(An alternative choice would be a log-normal distribution, which might
be more consistent with the expectation that for a natural distribution
values like 1/3 and 3 are equally likely.  Exploring the sensitivity
to choices of priors is part of the Bayesian analysis but is beyond
the scope of the present discussion, see~\cite{Furnstahl:2014aa}.)
With this gaussian form at fixed $M$ and $R$, 
the prior pdf combines with the likelihood pdf to define 
an \emph{augmented} $\chi^2$ function, which can be analytically minimized to 
find the maximum of the posterior pdf, with the corresponding uncertainties 
obtained from the least-squares analysis.
Typical results are shown in the right half of Table~\ref{tab:toy_results1}
(in this case for $R=5$).
Now we find the coefficients are stabilized with respect to the order $M$
and the breakdown of the fit is simply signaled by the values
of a coefficient being overwhelmed by the uncertainty. 

Suppose now that we have developed our EFT to order $p$, but we want to
account for missing contributions up to order $M > p$.
In the Bayesian approach at fixed $M$ and $R$, we treat the higher-order coefficients 
as ``nuisance parameters'' and marginalize over them to obtain the best estimate of 
$a_0$, $a_1$, $\ldots$, $a_p$, given the data and the expectation that 
\emph{all} the $a_i$'s are natural---see Eq.~(\ref{eq:marg}).
In the case of gaussian priors, these integrals
 can be done 
analytically (see Ref.~\cite{Schindler:2008fh}). 
The resulting probability 
distribution can \emph{in this case} also be understood as being associated with a modified, 
augmented $\chi^2$ function~\cite{Stump:2001gu}:
\bea
  \chi^2_{\rm aug} &=& \sum_{k=1}^{N}
    \left( \frac{d_k - \sum_{i=0}^p a_i (x_k)^i}{\sigma_k} \right)^2 
      - \sum_{j,j'=1}^{M-p} B_j  (A^{-1})_{j j'}  B_{j'} 
       + \sum_{i=0}^p \frac{a_i^2}{R^2}
       \;,
       \label{eq:chisqwithEFTerrors}
\eea
with the $(M-p)$--dimensional vector ${\bf B}$ and $M-p \times M-p$ matrix ${\bf A}$ defined by:
\begin{equation}
    B_j=\sum_{k=1}^N x_k^{j+p} \frac{d_k - \sum_{i=0}^M a_i (x_k)^i}{\sigma_k^2}; 
    \quad 
    A_{j j'}=\frac{\delta_{jj'}}{R^2} 
    + \sum_{k=1}^N \frac{x_k^{j + j' + 2 p}}{\sigma_k^2}
    \;.
    \label{eq:BjAjjp}
\end{equation}
We emphasize that this is {\it not} the same as adding a theoretical uncertainty
in quadrature to $\sigma_k$ (cf.\ Ref.~\cite{Dobaczewski:2014jga}).  This suggests that parameterizing the EFT Hamiltonian error via a $p^n$ penalty (as in the $O(Q^3)$ fit of NN data in Ref.~\cite{Ekstrom:2013kea}) is too simplistic.
Instead, Eq.~(\ref{eq:chisqwithEFTerrors}) accounts for the higher-order terms in the EFT Hamiltonian as correlated systematic errors in the EFT fit.

\begin{figure}[tbh-]
 \begin{center}
 \includegraphics[width=0.65\columnwidth]{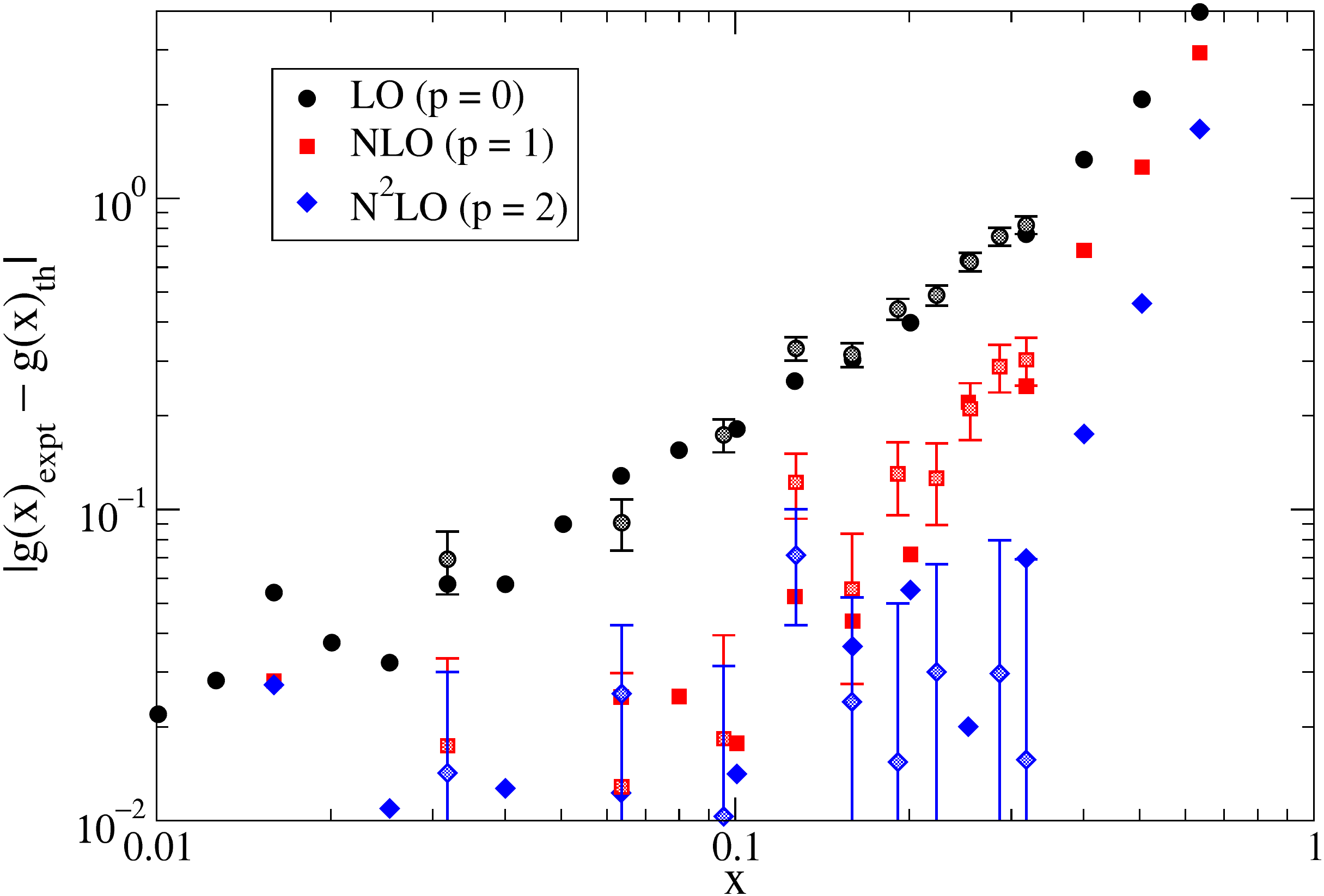}
 \caption{Residuals for $g(x)$ from the same synthetic data (``expt'') used in
 Table~\ref{tab:toy_results1} (with error bars showing its relative errors)
 plus some additional synthetic data covering a greater range in $x$, and
 the theoretical expectations $g(x)_{\rm th}$ for the first three orders
 of the Taylor series using the best estimates of $a_0$, $a_1$
 and $a_2$ from Eqs.~\eqref{eq:chisqwithEFTerrors} and \eqref{eq:BjAjjp}.
 }
 \label{fig:toylepage}
 \end{center}
\end{figure}

An appropriate error plot should reveal whether this marginalization has been
successful in implementing a systematic EFT expansion.
In Fig.~\ref{fig:toylepage}, the values for $a_0$, $a_1$
and $a_2$ from minimizing Eq.~\eqref{eq:chisqwithEFTerrors} 
are used for the theoretical estimates of $g(x)$ for $p=0$, $1$, and $2$ ($M=5$).
The residuals from synthetic data that determined
the coefficients are the points with error bars; the error-bar size reflects 
the 5\% relative error in the data.  
(Uncertainties in the coefficients are not included here.)
Residuals from additional synthetic data 
at a larger range of $x$ using the same coefficients help to show the
trends.  We see a clear power-law increase in the residuals at leading order ($p=0$),
where the theory error dominates the data errors. At low $x$ these residuals grow linearly, as expected.
For slightly larger $x$ the NLO ($p=1$) residuals are consistent with quadratic growth over a significant range of $x$.  Data errors ensure that the scatter of residuals
for $p=2$ obscures any power law at low and intermediate
$x$. The growth of the LO, NLO, and N$^2$LO residuals with $x$ is not consistent with the EFT order for high $x$, because the expansion converges only slowly there, but the systematic improvement and ultimate breakdown of the theory
is still manifest in this region. This simple exercise already makes it clear 
that definitive identification of
the power-law behavior of residuals is strongly dependent on 
the relative size of the statistical and theory errors.

\begin{figure}[t-]
 \begin{center}
  \includegraphics[width=0.32\columnwidth]{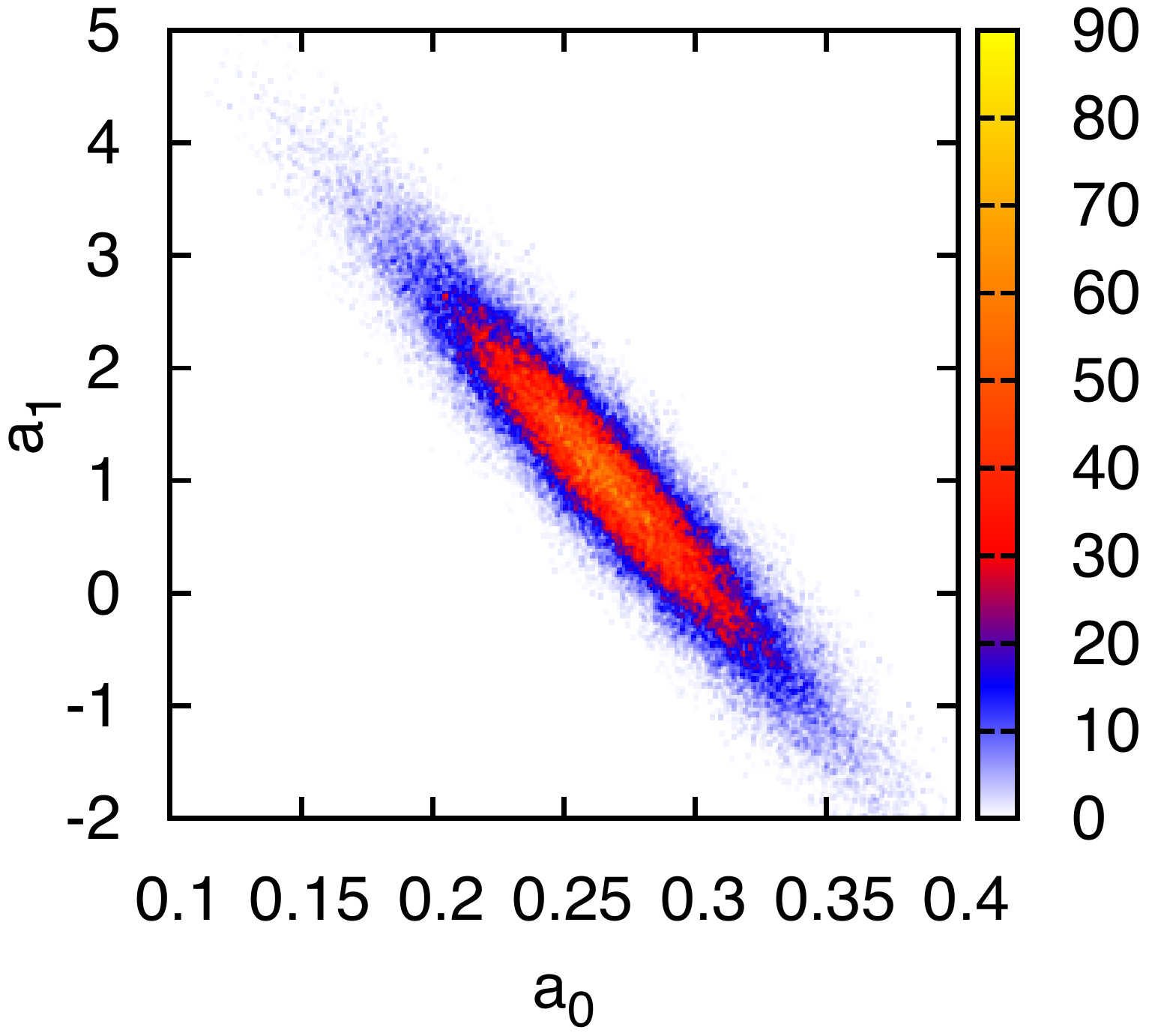}~~%
  \includegraphics[width=0.32\columnwidth]{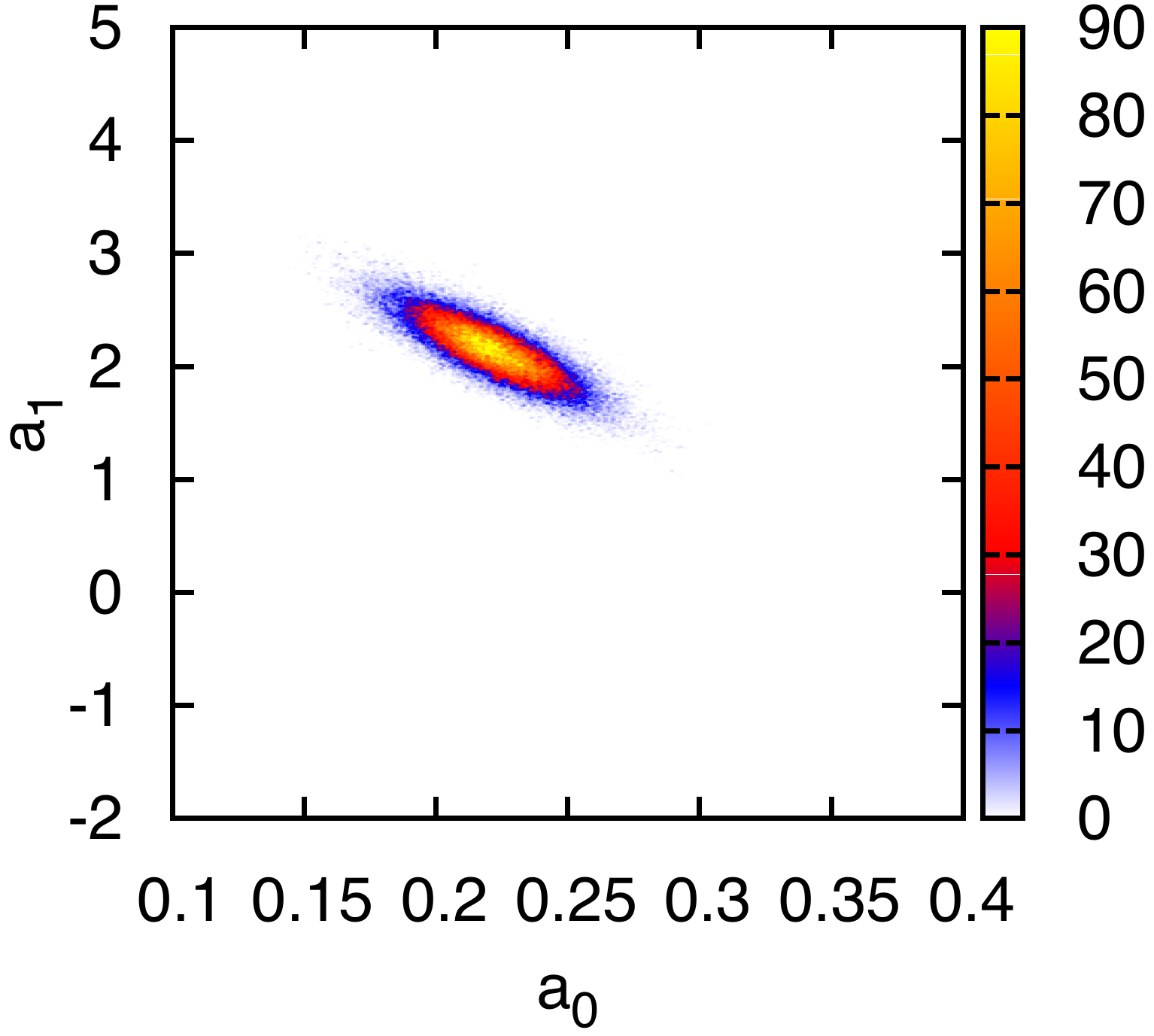}~~%
  \includegraphics[width=0.32\columnwidth]{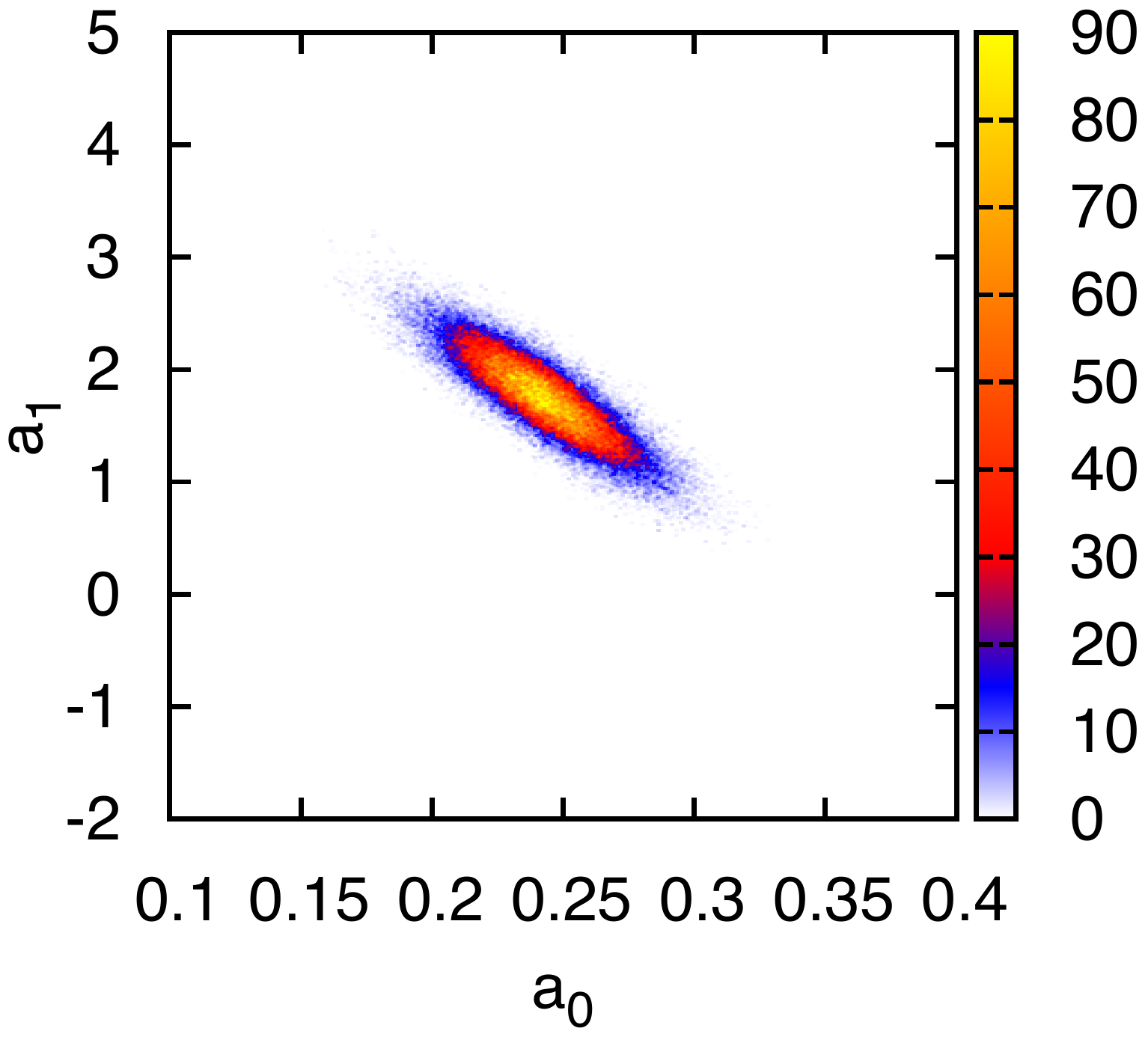}
 \caption{Projected posterior pdfs from MCMC sampling for two coefficients
 ($a_0$ and $a_1$) in the toy model with M=3. 
   a) No prior; b) gaussian prior with R=1; c) gaussian prior with R=5.
   The color scale indicates counts per histogram bin.} 
 \label{fig:posterior}
 \end{center}
\end{figure}

In more general cases we will not be able to do all of the integrals
in the expression for the posterior pdf analytically, and computing the posterior can become
a computationally difficult problem.  Fortunately, Markov-chain
Monte Carlo (MCMC) sampling methods, familiar from physics applications such as
statistical simulations or variational Monte Carlo, are readily adapted
to the problem~\cite{Sivia:06}. 
The output is an approximation to the posterior pdf for $\avec$, which
then can be analyzed according to the details of the problem.
In Fig.~\ref{fig:posterior}, we show examples of MCMC distributions for
$a_0$ and $a_1$ at fixed $M=3$ from the toy problem, first without a prior 
and then with a gaussian prior with two choices for $R$.
We emphasize that the full output from the Bayesian approach is a posterior
pdf for $\avec$ as a whole.
The plots illustrate the nature of the correlation between these parameters
and show how the addition of priors concentrates the posterior distributions, 
as already
implied by Table~\ref{tab:toy_results1}. 
The narrowness of a $\chi^2$ 
with many degrees of freedom means that the results can be quite sensitive to 
the choice of $R$~\cite{Stump:2001gu}; in Fig.~\ref{fig:posterior}
we see evidence that $R=1$ is too restrictive for a naturalness prior.
We can account for this to some extent by integrating over 
an appropriate pdf for $R$
(see Refs.~\cite{Schindler:2008fh,Furnstahl:2014aa}).

Thus far, we have only considered the determination and uncertainties
of the coefficients $\avec$, corresponding to the LECs in an EFT.
However, it is the observables and not the LECs that are of ultimate
interest. 
But the same probability distributions that appear in the computation of the 
posterior for an LEC appear in the predictive probability distribution for an 
observable of interest. 
Uncertainty intervals can then be determined using standard
Bayesian procedures~\cite{Sivia:06}.
Such predictions will automatically include not only 
the error bar from the data analysis, but also a component of the uncertainty 
from omitted higher-order terms in the EFT (Taylor-series) expansion. 
Examples of error propagation are given in Ref.~\cite{Furnstahl:2014aa}
along with generalizations of the toy problem that address other
issues in EFT uncertainty quantification.

\section{Summary and Outlook} \label{sec:summary}
  
Our recipe for EFT uncertainty quantification offers a vision of nuclear
calculations with statistically meaningful error bands on observables. Such a theory is
truly falsifiable, since it can make predictions, with error bars, that can be critically compared
with extant data or future experiments.
Furthermore, the tools we have outlined here permit identification of the contributions to the
error bands from different sources of uncertainty in the calculation.
This enables quantitative guidance regarding which
ingredient (experiment, many-body methods, EFT Hamiltonians) requires extra focus
if uncertainties are to be driven lower, thus enhancing
 the scientific cycle
between theory and experiment.

\section*{Acknowledgments}

We thank H.~Grie{\ss}hammer, D.~Higdon, A.~Steiner, A.~Thapaliya, and K.~Wendt  for useful discussions. We are grateful to H.~Grie{\ss}hammer and J.~A.~McGovern for preparing figures for this article, and to M.~Schindler for sharing the MATHEMATICA\textregistered~package BayesEFTfit with us.
This work was supported in part by the National Science Foundation
under Grant No.~PHY--1306250, the U.S. Department of Energy under 
grant DE-FG02-93ER40756, and the NUCLEI SciDAC Collaboration under
DOE Grant DE-SC0008533.

\section*{References}

\bibliographystyle{myunsrt}

\bibliography{bayesian_refs}

\begin{thebibliography}{10}

\bibitem{Beane:2000fx}
S.~R. Beane et~al.
\newblock From hadrons to nuclei: Crossing the border.
\newblock In M.~Shifman, editor, {\em At the Frontier of Particle Physics,
  Vol.~1}, pages 133--269. World Scientific, 2000.

\bibitem{Epelbaum:2008ga}
E.~Epelbaum, H.-W. Hammer, and U.-G. Meissner.
\newblock {Modern Theory of Nuclear Forces}.
\newblock {\em Rev. Mod. Phys.}, 81:1773--1825, 2009.

\bibitem{Bertulani:2002sz}
C.~A. Bertulani, H.-W. Hammer, and U.~Van~Kolck.
\newblock Effective field theory for halo nuclei.
\newblock {\em Nucl. Phys. A}, 712:37--58, 2002.

\bibitem{Papenbrock:2013cra}
T.~Papenbrock and H.~Weidenmueller.
\newblock {Effective Field Theory for Finite Systems with Spontaneously Broken
  Symmetry}.
\newblock {\em Phys. Rev. C}, 89:014334, 2014.

\bibitem{Machleidt:2011zz}
R.~Machleidt and D.~Entem.
\newblock {Chiral effective field theory and nuclear forces}.
\newblock {\em Phys. Rept.}, 503:1--75, 2011.

\bibitem{Griesshammer:2008aaa}
H.~W. Griesshammer.
\newblock Effective field theories in few-nucleon systems, 2008.
\newblock
  https://www.phys.gwu.edu/NNPSS/talks-speakers/griesshammer/griesshammer.pdf.

\bibitem{Schindler:2008fh}
M.~R. Schindler and D.~R. Phillips.
\newblock {Bayesian Methods for Parameter Estimation in Effective Field
  Theories}.
\newblock {\em Annals Phys.}, 324:682--708, 2009.

\bibitem{Lepage:1997cs}
G.~P. Lepage.
\newblock {How to renormalize the Schrodinger equation}.
\newblock 1997.
\newblock arXiv:1406.0625.

\bibitem{Griesshammer:2004pe}
H.~W. Griesshammer.
\newblock {Improved convergence in the three-nucleon system at very low
  energies}.
\newblock {\em Nucl. Phys. A}, 744:192--226, 2004.

\bibitem{Baru:2010xn}
V.~Baru et~al.
\newblock {Precision calculation of the $\pi^-$-deuteron scattering length and
  its impact on threshold $\pi$-N scattering}.
\newblock {\em Phys. Lett. B}, 694:473--477, 2011.

\bibitem{Weinberg:1990rz}
S.~Weinberg.
\newblock Nuclear forces from chiral lagrangians.
\newblock {\em Phys. Lett. B}, 251:288--292, 1990.

\bibitem{Ordonez:1995rz}
C.~Ordonez, L.~Ray, and U.~van Kolck.
\newblock The two-nucleon potential from chiral lagrangians.
\newblock {\em Phys. Rev. C}, 53:2086--2105, 1996.

\bibitem{Epelbaum:1999dj}
E.~Epelbaum, W.~Gl{\"o}ckle, and U.-G. Meissner.
\newblock {Nuclear forces from chiral Lagrangians using the method of unitary
  transformation}.
\newblock {\em Nucl. Phys. A}, 671:295--331, 2000.

\bibitem{Entem:2001cg}
D.~R. Entem and R.~Machleidt.
\newblock {Accurate nucleon-nucleon potential based upon chiral perturbation
  theory}.
\newblock {\em Phys. Lett. B}, 524:93--98, 2002.

\bibitem{Entem:2003ft}
D.~R. Entem and R.~Machleidt.
\newblock Accurate charge-dependent nucleon-nucleon potential at fourth order
  of chiral perturbation theory.
\newblock {\em Phys. Rev. C}, 68:041001, 2003.

\bibitem{Epelbaum:2004fk}
E.~Epelbaum, W.~Glockle, and U.-G. Meissner.
\newblock The two-nucleon system at next-to-next-to-next-to-leading order.
\newblock {\em Nucl. Phys. A}, 747:362--424, 2005.

\bibitem{VanKolck:1994yi}
U.~van Kolck.
\newblock Few nucleon forces from chiral lagrangians.
\newblock {\em Phys. Rev. C}, 49:2932--2941, 1994.

\bibitem{Epelbaum:2002vt}
E.~Epelbaum, A.~Nogga, W.~Gloeckle, H.~Kamada, U.-G. Meissner, and H.~Witala.
\newblock {Three-nucleon forces from chiral effective field theory}.
\newblock {\em Phys. Rev. C}, 66:064001, 2002.

\bibitem{Marji:2013uia}
E.~Marji, A.~Canul, Q.~MacPherson, R.~Winzer, C.~Zeoli, et~al.
\newblock {Nonperturbative renormalization of the chiral nucleon-nucleon
  interaction up to next-to-next-to-leading order}.
\newblock {\em Phys. Rev. C}, 88(5):054002, 2013.

\bibitem{Phillips:2013fia}
D.~R. Phillips.
\newblock {Recent results in chiral effective field theory for the NN system}.
\newblock {\em PoS}, CD12:013, 2013.

\bibitem{Bogner:2009bt}
S.~K. Bogner, R.~J. Furnstahl, and A.~Schwenk.
\newblock {From low-momentum interactions to nuclear structure}.
\newblock {\em Prog. Part. Nucl. Phys.}, 65:94--147, 2010.

\bibitem{Furnstahl:2013oba}
R.~J. Furnstahl and K.~Hebeler.
\newblock {New applications of renormalization group methods in nuclear
  physics}.
\newblock {\em Rept. Prog. Phys.}, 76:126301, 2013.

\bibitem{Stetcu:2006ey}
I.~Stetcu, B.~R. Barrett, and U.~van Kolck.
\newblock No-core shell model in an effective-field-theory framework.
\newblock {\em Phys. Lett. B}, 653:358--362, 2007.

\bibitem{Lee:2008fa}
D.~Lee.
\newblock {Lattice simulations for few- and many-body systems}.
\newblock {\em Prog. Part. Nucl. Phys.}, 63:117--154, 2009.

\bibitem{Stoks:1992ja}
V.~G.~J. Stoks, R.~Timmermans, and J.~J. de~Swart.
\newblock {On the pion--nucleon coupling constant}.
\newblock {\em Phys. Rev. C}, 47:512--520, 1993.

\bibitem{Georgi:1992dw}
H.~Georgi.
\newblock {Generalized dimensional analysis}.
\newblock {\em Phys. Lett. B}, 298:187--189, 1993.

\bibitem{Epelbaum:2001fm}
E.~Epelbaum, U.-G. Meissner, W.~Gloeckle, and C.~Elster.
\newblock {Resonance saturation for four nucleon operators}.
\newblock {\em Phys. Rev. C}, 65:044001, 2002.

\bibitem{Perez:2014yla}
R.~Navarro~Perez, J.~Amaro, and E.~Ruiz~Arriola.
\newblock {Statistical error analysis for phenomenological nucleon-nucleon
  potentials}.
\newblock {\em Phys. Rev. C}, 89:064006, 2014.
\newblock arXiv:1404.0314.

\bibitem{Ekstrom:2013kea}
A.~Ekstr{\"o}m, G.~Baardsen, C.~ForssŽn, G.~Hagen, M.~Hjorth-Jensen, et~al.
\newblock {Optimized chiral nucleon-nucleon interaction at
  next-to-next-to-leading order}.
\newblock {\em Phys. Rev. Lett.}, 110:192502, 2013.

\bibitem{Beane:2004ra}
S.~Beane, M.~Malheiro, J.~McGovern, D.~R. Phillips, and U.~van Kolck.
\newblock {Compton scattering on the proton, neutron, and deuteron in chiral
  perturbation theory to O(Q**4)}.
\newblock {\em Nucl. Phys. A}, 747:311--361, 2005.

\bibitem{McGovern:2012ew}
J.~A. McGovern, D.~R. Phillips, and H.~W. Griesshammer.
\newblock {Compton scattering from the proton in an effective field theory with
  explicit Delta degrees of freedom}.
\newblock {\em Eur. Phys. J. A}, 49:12, 2013.

\bibitem{Entem:2002sf}
D.~Entem and R.~Machleidt.
\newblock {Chiral 2pi exchange at order four and peripheral NN scattering}.
\newblock {\em Phys. Rev. C}, 66:014002, 2002.

\bibitem{Rentmeester:2003mf}
M.~C.~M. Rentmeester, R.~G.~E. Timmermans, and J.~J. de~Swart.
\newblock {Determination of the chiral coupling constants $c_3$ and $c_4$ in
  new pp and np partial-wave analyses}.
\newblock {\em Phys. Rev. C}, 67:044001, 2003.

\bibitem{Phillips:1999hh}
D.~R. Phillips, G.~Rupak, and M.~J. Savage.
\newblock {Improving the convergence of NN effective field theory}.
\newblock {\em Phys. Lett. B}, 473:209--218, 2000.

\bibitem{Nogga:2004ab}
A.~Nogga, S.~K. Bogner, and A.~Schwenk.
\newblock Low-momentum interaction in few-nucleon systems.
\newblock {\em Phys. Rev. C}, 70:061002, 2004.

\bibitem{Gazit:2014}
D.~Gazit.
\newblock Talk given at the ECT* workshop on Three-nucleon Forces. ECT*,
  Trento, May 2014.

\bibitem{Sivia:06}
D.~Sivia and J.~Skilling.
\newblock {\em Data Analysis: A Bayesian Tutorial}.
\newblock Oxford University Press, 2006.

\bibitem{deFinetti:1974}
B.~de~Finetti.
\newblock {\em Theory of Probability: A Critical Introductory Treatment}.
\newblock Wiley, 1974.

\bibitem{Jaynes:1957}
E.~T. Jaynes.
\newblock {\em Phys. Rev.}, 106:620, 1957.

\bibitem{Dobaczewski:2014jga}
J.~Dobaczewski, W.~Nazarewicz, and P.-G. Reinhard.
\newblock {Error Estimates of Theoretical Models: a Guide}.
\newblock {\em J. Phys. G}, 41:074001, 2014.

\bibitem{Jeffreys:1939}
H.~Jeffreys.
\newblock {\em Theory of Probability}.
\newblock Clarendon Press, 1939.

\bibitem{Arina:2011si}
C.~Arina, J.~Hamann, and Y.~Y. Wong.
\newblock {A Bayesian view of the current status of dark matter direct
  searches}.
\newblock {\em JCAP}, 1109:022, 2011.

\bibitem{Pato:2012fw}
M.~Pato, L.~E. Strigari, R.~Trotta, and G.~Bertone.
\newblock {Taming astrophysical bias in direct dark matter searches}.
\newblock {\em JCAP}, 1302:041, 2013.

\bibitem{Rieping:2005}
W.~Rieping, M.~Habeck, and M.~Nilges.
\newblock {Inferential Structure Determination}.
\newblock {\em Science}, 309:303--306, 2005.

\bibitem{Lepage:2001ym}
G.~Lepage, B.~Clark, C.~Davies, K.~Hornbostel, P.~Mackenzie, et~al.
\newblock {Constrained curve fitting}.
\newblock {\em Nucl. Phys. Proc. Suppl.}, 106:12--20, 2002.

\bibitem{Cabrera:2008tj}
M.~Cabrera, J.~Casas, and R.~Ruiz~de Austri.
\newblock {Bayesian approach and Naturalness in MSSM analyses for the LHC}.
\newblock {\em JHEP}, 0903:075, 2009.

\bibitem{Ghilencea:2012qk}
D.~Ghilencea and G.~Ross.
\newblock {The fine-tuning cost of the likelihood in SUSY models}.
\newblock {\em Nucl. Phys. B}, 868:65--74, 2013.

\bibitem{Balazs:2013qva}
C.~Balazs, A.~Buckley, D.~Carter, B.~Farmer, and M.~White.
\newblock {Should we still believe in constrained supersymmetry?}
\newblock {\em Eur. Phys. J.}, C73:2563, 2013.

\bibitem{Cacciari:2011ze}
M.~Cacciari and N.~Houdeau.
\newblock {Meaningful characterisation of perturbative theoretical
  uncertainties}.
\newblock {\em JHEP}, 1109:039, 2011.

\bibitem{Forte:2013mda}
S.~Forte, A.~Isgrò, and G.~Vita.
\newblock {Do we need N$^3$LO Parton Distributions?}
\newblock {\em Phys. Lett. B}, 731:136--140, 2014.

\bibitem{landau2011survey}
R.~Landau, J.~P{\'a}ez, and C.~Bordeianu.
\newblock {\em A Survey of Computational Physics: Introductory Computational
  Science}.
\newblock Princeton University Press, 2011.

\bibitem{Bedaque:2002yg}
P.~F. Bedaque, G.~Rupak, H.~W. Griesshammer, and H.-W. Hammer.
\newblock {Low-energy expansion in the three-body system to all orders and the
  triton channel}.
\newblock {\em Nucl. Phys. A}, 714:589--610, 2003.

\bibitem{Furnstahl:2014aa}
R.~J. Furnstahl, N.~Klco, D.~R. Phillips, A.~Thapaliya, and S.~Wesolowski.
\newblock 2014.
\newblock In preparation.

\bibitem{Stump:2001gu}
D.~Stump et~al.
\newblock {Uncertainties of predictions from parton distribution functions. 1.
  The Lagrange multiplier method}.
\newblock {\em Phys. Rev. D}, 65:014012, 2001.

\end{thebibliography}

\end{document}